# A Review on Dyadic Conversation Visualizations – Purposes, Data, Lens of Analysis

Joshua Y. Kim, Rafael A Calvo, Kalina Yacef, N.J. Enfield

**Abstract**— Many professional services are provided through text and voice systems, from voice calls over the internet to messaging and emails. There is a growing need for both individuals and organizations to understand these online conversations better and find actionable insights. One method that allows the user to explore insights is to build intuitive and rich visualizations that illustrate the content of the conversation. In this paper, we present a systematic survey of the various methods of visualizing a conversation and research papers involving interactive visualizations and human participants. Findings from the survey show that there have been attempts to visualize most, if not all, of the types of conversation that are taking place digitally – from speech to messages and emails. Through this survey, we make two contributions. One, we summarize the current practices in the domain of visualizing dyadic conversations. Two, we provide suggestions for future dialogue visualization research.

## 1 Introduction

A wide range of rich information is exchanged when people communicate. At the simplest level, simple informational exchange refers to the meaning of what is being spoken or written. Underneath the surface form of the utterance, there could be rhetorical aspects that are latent in nature [1]. To add to the complexity, the prosody of a speech and the aesthetics of a handwritten document give us hints about the emotions and attitudes of the speaker or author at the time of speaking or writing. In short, to be skilled in written and spoken communication, one must – consciously or not – keep track of a wide range of information including content, themes, emotions, attitudes, power or cultural differences and many more.

In the digital age, a significant portion of our communication now is technology-mediated. As companies increasingly use digital channels to communicate with their customers, there is an increasing number of commercial tools that facilitate such conversations. Public relations professionals are increasingly using digital channels like Twitter to broadcast and initiate one-to-one synchronous (for example, live pop-up chat to improve website conversions like Intercom [2]) or asynchronous conversations (for example, personalized emails using tools like Salesforce [3]). With the increase in online conversations, a summary of such conversations in the form of visualization could be useful, yet our review suggests that there have been limited studies on methods to visualize dyadic conversations.

On the one hand, we need new ways we understand and summarize conversations to improve productivity or augment human cognition. On the other hand, such techniques can also help improve human-human communication by helping develop the professional's communication skills. Both improving efficiency through better visual 'summaries' and improving human-human communication skills are important in many domains: health and counseling, teaching, help-desk, for example.

Good visualizations of the content of a conversation facilitate the understanding of its structure and information content, thereby augmenting human cognition. Given the potential of visualizations to augment human cognition in a world of ubiquitous online conversations, we aim to contribute by helping researchers gain an understanding of the state-of-the-art in dyadic conversation visualizations.

Dyadic conversations are particularly relevant and are the focus of this review. The exclusion of group conversations and focus on dyadic conversations is not only due to its specific applications but also because dyads are qualitatively different from groups and those insights gathered from studying dyads will not always apply to groups [4].

The paper addresses the need for conceptual and practical findings of visualizations that can be used to understand online dyadic conversations. The paper is structured as follows: Section 2 describes the methodology used to identify relevant literature. Section 3 discusses the purposes for the visualization designs and the methods to measure the success of the visualizations. Section 4 introduces the types of textual data used in the studies included in this paper. Section 5 discusses the lenses of analysis through which conversational content have been visualized in the included studies. Finally, we conclude this paper with a discussion of the current gaps and future

- *Joshua Y. Kim is with the University of Sydney, School of Computer Science, Australia. E-mail: josh.kim@sydney.edu.au.*
- *Rafael A. Calvo (corresponding author) is with the University of Sydney, School of Electrical and Information Engineering, Australia. E-mail: rafael.calvo@sydney.edu.au.*
- *Kalina Yacef is with the University of Sydney, School of Computer Science, Australia. E-mail: kalina.yacef@sydney.edu.au.*
- *N.J. Enfield is with the University of Sydney, Faculty of Arts and Social Sciences, Australia. E-mail: nick.enfield@sydney.edu.au.*

opportunities in this domain.

## 2 Survey methods

The literature in visualizations of language is highly multidisciplinary and often uses mixed-methods. Our review of the literature was conducted using the mixed methods research synthesis framework (MMRS) [5] – developed to combine qualitative and quantitative primary studies – over the past 20 years and across disciplines.

In this review, our objective is to answer the question, "What computer-generated visualizations have been built to visualize the content of a conversation between two people, for what purpose and what were their significant features?" We considered a 20-year window in this review, from 1998 to 2018, and performed multiple searches using ACM, IEEE, Web of Science, and Scopus. The process is iterative, and the search terms widen as new keywords and database indices emerge. The search terms are provided in Appendix 3, and the last search was performed in December 2018.

There are three main aspects covered in this review: what are the purposes of building the visualization and how to measure its success (Purposes), the data type at hand (Data), what metaphor has been employed before to analyze the content (Lens of analysis). Following the MMRS framework, we start by organizing the articles in the context of *Purposes*, *Data*, *Lens of analysis* through thematic analysis. Then, we provide a brief quantitative synthesis of user studies performed in the included studies.

In *Purposes*, we aim to address the question, "what was the purpose of building the visualization?" We then discuss the different methods through which success is measured from different aspect; it could be user performance, user experience or visualization algorithms efficiency.

In *Data*, we focus exclusively on textual data and exclude articles that focus solely on non-verbal communication such as gaze, speech rate or prosody without considering the semantic content of the text. For voice conversations, the study would be included if the voice conversations were transcribed into text. In addition, since we focus on dyadic conversations, the study must describe at least one visualization of a conversation between two people to be included. This means studies that only visualize the content of monologues – news and research articles for example – would be excluded. Studies that only investigate the conversation between a human and a bot would also be excluded. Lastly, studies that only visualize group conversations including more than two people would be excluded.

In *Lens of analysis*, we focus on the metaphors that visualize the content of the conversation. Studies that visualize only the meta-data of the communication, such as counting the number of emails by the sender or time of day are excluded. As for visualizations that involve visualizing emotions, we only included studies that automatically extract the emotions content and excluded those that only allow the user to manually specify an emotional expression to be communicated. This distinction ensures that the scope does not creep out to include visualizations that are manually supplemented by the user. Similarly, only visualizations that automatically extract the argument structure of debates are included.

### 2.1 Overview of studies selection

A total of 3,357 articles were retrieved from the following databases: ACM DL (N=3085), IEEE Xplore (N=165), Scopus (N=65), and Web of Science (N=42). Fig. 1 shows an overview of the process. "Additional records identified through other sources" come from the review of reference lists of included studies. Using DistillerSR, duplicates are automatically identified and removed, returning 3113 distinct articles. The title and abstract screen removed 2952 articles, while the full-text screen removed a further 135 articles with its reason and associated count depicted in the figure. At the end of both screens, 26 articles are included in this review.

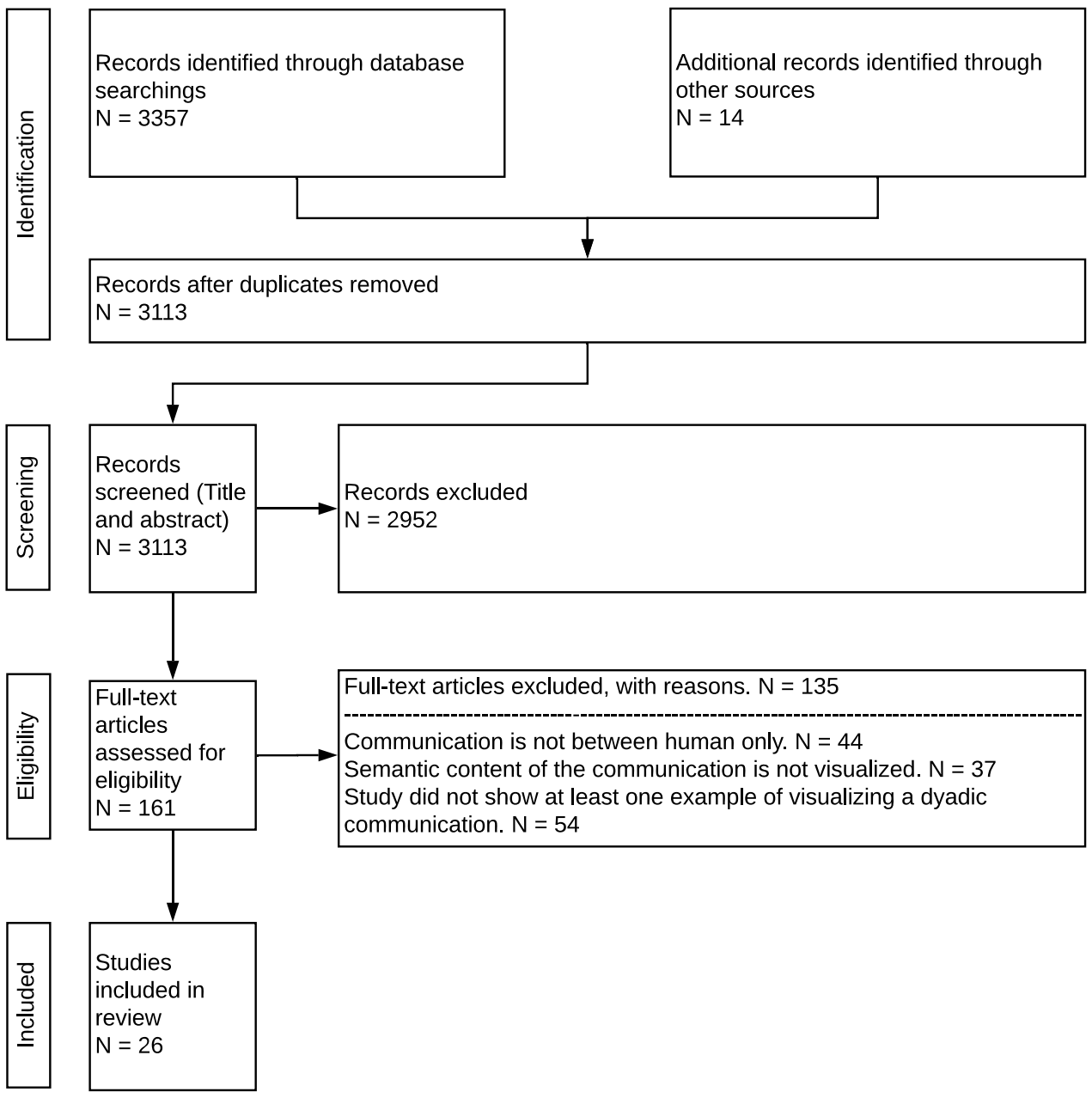

Fig. 1 – Flow of articles through different phases of the systematic review

Table 1 provides an overview of the included papers, and the definition of each column is as follow. Lens of analysis refers to the type of content being visualized. Data refers to the specific type of textual data used in the study. The Study Type column is closely tied to the purposes for the visualizations – researchers may have an intended outcome in mind when designing a visualization, and therefore may design a qualitative (Qual.) or quantitative (Quant.) experiment to test for this intended outcome. SR Index refers to the index number of the included study which would be used to refer to the study for the rest of this review. Fig. index refers to the figure index number for the corresponding visualization screenshot for each included study in Appendix 2.

Table 1
Overview of papers that visualized the content of a dyadic conversation using textual data.

| Lens of analysis | Data | Reference | Study Type (Sample Size) | SR Index | Fig. Index |
|---|---|---|---|---|---|
| Words and Phrases | (E) | [6] | - | 08 | Fig. 6 |
| | | [7] | Both (6) | 16 | Fig. 14 |
| | | [8] | Qual. (16) | 24 | Fig. 5 |
| | (I) | [9] | - | 05 | Fig. 3 |
| | (S) | [10] | - | 13 | Fig. 4 |
| | | [11] | - | 17 | Fig. 2 |
| | | [12] | - | 19 | Fig. 10 |
| | | [13] | - | 25 | Fig. 7 |
| Themes | (E) | [14] | - | 11 | Fig. 16 |
| | (I) | [15] | Qual. (7) | 12 | Fig. 15 |
| | (M) | [16] | Both (7) | 14 | Fig. 11 |
| | (S) | [17] | Quant. (48) | 04 | Fig. 17 |
| | | [18] | - | 06 | Fig. 8,9 |
| | | [19] | Qual. (3) | 15 | Fig. 12 |
| | | [20] | - | 18 | Fig. 13 |
| | | [21] | - | 21 | - |
| | | [22] | - | 22 | - |
| | | [23] | Quant. (9) | 26 | Fig. 26 |
| Argument Structure | (I) | [24] | Qual. (10 to 20) | 02 | Fig. 19 |
| | | [25] | - | 10 | Fig. 18 |
| | | [26] | Both (16) | 23 | Fig. 20 |
| Emotions | (E) | [27] | Both (20) | 01 | Fig. 23 |
| | (I) | [28] | Quant. (10) | 03 | Fig. 25 |
| | | [29] | Qual. (Several) | 07 | Fig. 21 |
| | | [30] | Quant. (20) | 20 | Fig. 24 |
| | (M) | [31] | Qual. (6) | 09 | Fig. 22 |

Note (E): Email, (I): Instant Messaging, (M): Mobile Messaging, (S): Speech Transcripts

### 2.2 Dyadic human-human conversations

In this section, we discuss the reasons for focusing on visualizations that analyze dyadic human-human conversations. Fig. 1 shows that 54 articles are excluded because the visualization did not visualize dyadic conversations, and a further 44 articles excluded because it involved a bot. Whilst excluding these articles, we appreciate that there are notable works in the realm of visualizing group conversations that are not included because of this decision. For example, group conversation visualizations in the online chat setting [32], [33]; in-person meetings [34]–[38]; asynchronous online group conversations [39], [40]; public reasoning [41]. In the next few paragraphs, we explain why dyadic conversations should not be deemed simply as a subset of group conversations where the number of participants equals to two. Dyadic conversations are qualitatively different in nature; thereby motivating the reasons for focusing on dyadic human-human conversations visualizations.

As discussed by Moreland [4], dyadic conversations raise particular challenges– (i) people feel strong and different emotions in dyads than in groups and (ii) some group phenomena does not occur in dyads. A relatable anecdotal example of how the range and degree of emotions are different would be to contrast a conversation between a romantic couple and a group conversation between four people, made up of two romantic couples. In group settings, social norms are developed which weakens emotional experiences of its members [42], which is a plausible explanation to why people enjoy smaller groups than larger ones, and dyads more so than small groups [43]. By limiting our analysis to dyadic conversations, we focus on the distinctive repertoire of emotions that are prevalent in dyadic settings.

Secondly, the dynamics of group conversations are very different from those between dyads. One aspect of the difference comes from turn-taking patterns, group discussions are often like monologues, and members are influenced by the dominant speaker [44]. When the typical group conversation is skewed towards one speaker, there might be different motivations when visualizing one-speaker dominant group conversations as compared to the more equally-distributed dyadic conversations. In addition, the benefit of focusing on dyadic conversation also means that we are limiting the dimensionality of the number of participants. Therefore, more dimensions (such as space and colors) can be afforded to visualizing the content of the conversation, thereby maximizing the richness of content visualizations.

Lastly, the nature of human-to-bot conversation can also be different from human-to-human conversations. In a general, unconstrained context, current dialogue systems are incapable of generating responses that are rated as highly appropriate by humans [45]. In addition, Hill et al. [46] found that when the participants are aware that they are speaking with a bot, they speak differently – with shorter messages, less rich vocabulary and greater profanity. Given the differences between human-to-bot conversations and human-to-human conversations, we elected to focus on human-human conversations for this paper.

## 3 Purposes

### 3.1 Overview of purposes

In this section, we summarize the purposes by extracting groups of target users from the included studies. We observe that there are three distinct groups of target users – regular users, independent researchers, and expert users. The regular user is personified as the common person who is in the conversation. The independent researcher is personified as the third-party analyst who is analyzing the conversation. The expert user is personified as the expert in the conversation, for example, the doctor or psychologist in the consulting conversation. Having defined these groups, we observe that these different groups focused on different elements within the conversation(s).

Broadly speaking, the different groups focused on three different elements, emotions, cognitive (temporal) and cognitive (non-temporal). Emotions refer to the emotional content within the conversation, whilst cognitive content refers to the non-emotional content. We further subdivide the cognitive group into temporal and non-temporal. Temporal visualizations include a time-element in the metaphor, to facilitate comparisons over time. Non-temporal visualizations are snapshots taken at the end of the conversation(s).

Table 2
Overview of the user groups and whether emotional content and temporal elements are included in each of the study.

| User Group | Cognitive Non-Temporal | Temporal | Emotion Both |
|---|---|---|---|
| Regular | SR02 | SR11 | SR01 |
| | SR10 | SR12 | SR03 |
| | SR23 | SR16 | SR05 |
| | | SR24 | SR07 |
| | | SR25 | SR09 |
| | | | SR20 |
| Independent Researcher | SR04 | SR06 | |
| | SR13 | SR08 | |
| | SR17 | SR19 | |
| | SR18 | SR21 | |
| | | SR22 | |
| Expert | SR15 | SR14 | |
| | | SR26 | |

Table 2 illustrates that most of the studies are for regular and independent researchers, with only three studies geared towards expert users.

For regular users, we would first focus on non-temporal visualizations. Debates and emotions visualizations are two relatively more significant areas of research for regular users, accounting for 9 out of 14 included studies. The other five studies belong to the temporal category, where topics are visualized over time.

For independent researchers, we see a wider range of application in non-temporal visualizations (SR04, SR13, SR17, and SR18). Two of these studies (SR04 and SR13) provided a snapshot of the semantic content of the conversation. SR17 provided an understanding of how words contribute towards an automatic machine classifier. Lastly, SR18 provided a comparison of speech-act profiles between two speakers using a radar-chart.

As for temporal visualizations aimed at independent researchers, recurrence plots that facilitate analysis of talk turns (SR06, SR19, SR21, SR22) account for four out of five included studies. The remaining study (SR08) places word clouds across the X-axis which represent time, allowing the user to observe the evolution of content over time.

Regarding the expert users, we have two observations. Firstly, the expert users are from the health-care sector such as psychology counselors (SR14 & SR26) or doctors (SR15). There are other conversations that include expert opinion – such as lawyers and accountants in the business domains. However, there are no visualizations made in these domains yet. Secondly, we note that none of the three visualizations within the health-care sector visualized the emotional content of the conversations. Instead, emotion visualizations are aimed exclusively at the regular user.

## 3.2 Measuring the success of the visualizations

In this section, we present the methods through which the success of the visualizations has be evaluated. Lam et. el. [47] provided a framework that is useful for organizing the three subsections in this section – evaluating user performance, evaluating user experience and automated evaluation of visualizations. Table 3 shows the distribution of included studies across the different modes of evaluation.

Table 3
Distribution of included studies across modes of evaluation

| Mode of evaluation | Included studies |
|---|---|
| Evaluating user performance | SR04 |
| Evaluating user experience | SR01, SR02, SR03, SR07, SR09, SR12, SR14, SR15, SR16, SR20, SR24, SR26 |
| Evaluating both user performance and experience | SR23 |

Evaluating user performance refers to the measurement of objectively measurable metrics such as time and error rate, whilst evaluating user experience refers to the seeking of people' subjective feedback and opinions [47]. To evaluate user performance and user experience, user studies are one way to test whether the visualization has been successful in achieving its objectives. However, of the studies included, only about half of the studies (14 out of 26) included some form of user study – quantitative or qualitative. For 7 out of the 26 studies, the authors had an outcome (hypothesis) in mind and designed a quantitative user study to test the hypothesis. Generally, we found that the typical sample size is about ten to fifteen subjects as seen in Table 1, which is in line with the recommendation from Faulkner [48] that with ten users, the coverage of usability issues is about 80%.

### 3.2.1 Evaluating user performance

After interacting with the visualization, the researcher evaluates the user performance by analyzing the speed and accuracy of the response given to the specific task involved. For example, SR04 uses measurements taken while the user is answering questions, such as the time taken per question and the accuracy of answers provided. These measurements are compared to the baseline performance where the user is also exposed to another variant of the visualization.

### 3.2.2 Evaluating user experience

The researcher might also evaluate the user's experience by asking the participants to score the effectiveness of the visualizations. Whilst quantitative measures differ widely by the type of content that is visualized, they generally include some form of scores that measures usefulness and satisfaction. None of the included studies employed the use of a standardized instrument like the System Usability Scale (SUS) [49], which might be useful when the results from a study could be benchmarked against a large number of SUS scores as compiled by Bangor, Kortum, & Miller [50]. They have also compiled a list of alternative non-proprietary standardized instruments which might be helpful for the prospective visualization researcher.

In evaluating user experience in emotional visualization studies, the performance metrics included how it made the users feel. Three studies visualized the emotional content and they all included a quantitative user study (SR01, SR03 & SR20). These three studies used an avatar to visualize the emotional content, and we analyzed the list of 21 questions. There are five main types of measurements, namely – accuracy of emotional content extraction, sense of co-presence, entertainment, system interactivity, and overall score. In addition, a key finding that emerges from the three studies is that such visualizations do not necessarily improve a sense of co-presence or entertainment.

#### *3.2.3 Automated evaluation of visualizations*

Lastly, the discussion on automated evaluation of visualizations is minimal in the included studies. Automated evaluation of visualizations is concerned with machine-related ratings, such as the speed at which the visualization is rendered. We observed that the included studies are multidisciplinary in nature, instead of being algorithmic-centric, which could be the reason why we did not find any papers that discussed automated evaluation at length.

## 4 Data

This section describes the three types of data used in the included studies – speech transcripts, instant/mobile messaging, and emails. These three types of communication differ in many aspects, for example, length, duration, purpose, and formality, which may have implications on the visualization design.

### 4.1 Speech Transcripts

Speech is the most natural form of communication between humans and is a form of multi-modal communication. The raw speech needs to be recorded and transcribed into text before it can be used for analysis. In our review, we found that most researchers used private proprietary datasets and that only SR18 and SR04 used a high quality and large dataset that is publicly available. Because we observed scarcity of publicly available datasets in the reviewed studies, we recommend the review of Serban et el. [51] for a list of publicly available transcription corpus.

Compared to emails and instant/mobile messaging, speech transcripts are unique, because:

- The synchronous nature of speech conversation puts time pressure on the speakers – Stivers et al. [52] found that most talk turn transitions occur within 0.2 seconds. This social norm of quick transitions encourages the participants sometimes to value the time to reply more than the grammatical correctness of the utterance.
- Coreference (when two or more expressions in a text denote the same referent) issues are made more complicated for the analysis if the speaker is speaking and pointing at the reference object as he speaks. This type of multi-modal expression is not captured in the transcript, thereby missing some contextual information.
- As each utterance is ephemeral, the listener may seek repair actions for the original speaker to repeat or clarify what is being said. There are two broad types of repair actions – weak (when the listener says merely "huh?" or equivalent, which effectively requests the original speaker to repeat the utterance) or strong (when the listener repeats or paraphrases what is being heard to seek confirmation) [53], [54].
- The prosody and facial expression of the speaker has a strong influence on how the words are to be interpreted; for example, detecting sarcasm using only text data has proven to be difficult [55]. In the studies that visualized the content of speech conversations, prosody was not considered together with the text data. We posit that this is an area for improvement given that prosody has a heavy influence on both the interpretation of the words being said and the emotional state of the speaker.
- Particularly in interviews, talk turns in speech transcripts often refer to earlier points or questions that were asked many talk-turns ago. Interviews are typically not conducted over instant/mobile messaging nor emails.

### 4.2 Instant and mobile messaging

Instant messaging refers to synchronous chat where messages are exchanged, and interlocutors typically expect a prompt reply. Mobile messaging refers to the short text message sent over the Short Message Service (SMS); such messages are typically under 160 characters. Increasingly, the line between instant and mobile messaging is becoming blurred in multi-platform applications like Whatsapp, Facebook Messenger, and WeChat. Hence, it makes sense to discuss their distinctive features – relative to emails and speech transcripts – as one:

- Some instant messengers indicate whether the other party is online or typing, which helps the user to set expectations on whether a reply is coming soon. Conversations may end abruptly as one of the parties attends to something else.
- Emoticons and animated gifs are widely used in this form of communication to express emotions and sentiments. In addition, terms like "lol and haha" are also commonly used to inject emotions and sentiments.
- Informal abbreviations are common, e.g., "idk (I don't know), lol (laugh out loud)".
- Unlike speech transcripts, instant and mobile messaging (as well as email) users are likely to hold multiple concurrent conversations. Conversation switching is therefore common, and it adds to the cognitive load of users.

### 4.3 Emails

Emails can be exchanged between two or more parties. For this review, we are focusing on studies that have visualized an email conversation between two people. Like the previous sections, we discuss the distinct features of emails:

- Emails typically are long form and are typically longer than each utterance in an instant messaging or mobile messaging setting.
- Emails do not show whether the author is online

or not; thus, the recipient does not have expectations of when the reply is coming, or if there will be any reply at all.

- Emails are used in both formal and informal settings. In formal settings, the grammatical structure is likely to be better than informal settings.
- While the subject line is helpful in identifying the topic, it is common for email conversations to drift away from the original subject, without changing the original subject line indicated on the email.
- Unlike speech transcripts – where it is expensive to transcribe all conversations with a single person into text – multiple email conversations with a single email address can be concatenated to provide a long-running history, making multi-year analysis possible.

## 5 Lens of analysis

The content of dyadic conversations is analyzed through four lenses: (1) Words and phrases, (2) Themes, (3) Argument Structures, and (4) Emotions. Table 4 provides a list describing each of the four lenses of analysis.

Table 4
List of the four lenses of analysis

| Lens of analysis | Description | Included studies |
|---|---|---|
| Words and phrases | Visualizations that include the display of words and phrases from the conversation in its metaphor. | SR05, SR08, SR13, SR16, SR17, SR24, SR25 |
| Themes | Visualizations that allow its users to inspect underlying themes at the message-level or conversation-level. | SR04, SR06, SR08, SR11, SR12, SR13, SR14, SR15, SR16, SR18, SR19, SR21, SR22, SR26 |
| Argument structures | Visualizations that summarizes the interaction of arguments in a debate. | SR02, SR10, SR23 |
| Emotions | Visualizations that visualize the automatically extracted emotions. | SR01, SR03, SR07, SR09, SR20 |

In this section, we focus on the presentation of the visualizations, and the purposes of building the visualizations would be discussed in the next section.

### 5.1 Words and phrases

Studies had different rationales for visualizing exact words and phrases in a conversation. Firstly, one might want to visualize the contribution of each word towards a classification decision on an utterance (Fig. 14 from SR16, Fig. 2 from SR17). Secondly, having an overview of the words used gives an idea of what the conversation is about (Fig. 3 from SR05, Fig. 4 from SR13, Fig. 5 from SR24, Fig. 6 from SR08). Lastly, having the words on a screen helps the speaker remember what is being said in a voice conversation (Fig. 7 from SR25) – as discussed previously, in Section 4.1 Speech transcripts, a voice conversation is ephemeral.

There are a few approaches to visualize this type of content, ranging from stylizing the font face to a more complex animation. Most visualizations in this subsection do not contain animations. We begin our discussion with the simplest form – the highlighting of words. In two studies (Fig. 2 from SR17 and Fig. 14 from SR16), the highlighting of words represents their contribution towards a classification decision. Specifically, in Fig. 14 from SR16, users have a multi-faceted view to browse the emails. In one window, a user could click on a theme that he/she is interested in, and the keywords belonging to the selected theme in the email would be highlighted. Similarly, in Fig. 2 from SR17, the objective is to classify each utterance in a conversation to a dialogue-act, and the visualization is created to illustrate which words were highly-weighted (important) in a classification model to perform dialogue act classification.

In addition to highlighting the words, other visualizations have changed the font size and the placement of the words to represent specific concepts. It is common to use different colors to represent different speakers (Fig. 3 from SR05, Fig. 7 from SR25). Placement of words in a spatial array can represent points in time (Fig. 3 from SR05, Fig. 5 from SR24, Fig. 7 from SR25) or clustering of concepts (Fig. 4 from SR13). It is also possible that both the color and placement of words do not have specific representation but only optimized for aesthetics (Fig. 6 from SR08).

### 5.2 Themes[1]

The theme of the conversation is sometimes pre-established in some studies and thus apparent. However, more often, themes are hidden within the conversation, so such visualizations are the most common content type amongst the included studies. To facilitate our discussion, we divide the visualizations into two groups (a) visualizing themes within a conversation and (b) visualizing themes across conversations.

#### *5.2.1 Themes within a conversation*

As a conversation develops, multiple themes might have been discussed, and it might be time-consuming to read the entire transcript to follow the various themes that were discussed. Visualizations discussed in this subsection help alleviate this problem by visualizing the themes that were discussed. There are four main types of metaphors being employed here – recurrence plots, stylized bar plots, radar charts and Gantt chart – which are discussed in turn.

Firstly, four studies used Discursis [56] to analyze speech transcripts in different professional settings. The salient visualization technique across all the four studies is the conceptual recurrence plot. Through making connections between talk turns that have similar themes, the conceptual recurrence plot is adept at illustrating talk turns that are referring to an earlier point made, or question asked in interviews. Like Discursis, Fig. 10 from SR19 illustrates how recurrence of syntactic bigrams could be visualized. Using the recurrence plot, the researcher could see the density of syntactic recurrence within some distance from a diagonal line of incidence. Recurrence plots facilitate quick comparisons across talk-turns within a conversation.

[1] Or topics, we use both terms interchangeably in this paper.

Secondly, whilst the design of recurrence plots is well-suited to a third-party analyst who tries to understand the conversation, in some use-cases, it is the speaker who requires real-time support to monitor the conversation. An example would be crisis counseling, Fig. 11 from SR14 designs a real-time visualization system that reduces counselor cognitive overload as the counselor engages in more than one concurrent session.

Thirdly, when considering studies that visualize the themes of a conversation, two studies employed radar charts to compare the profiles of the two speakers (Fig. 12 from SR15, Fig. 13 from SR18). The axes of the radar charts represent categories of each utterance, and the prevalence of each category determines the position of each point on the axes. In both studies, color is used to represent the different speakers, and the metaphor allows easy comparison of two user profiles.

Lastly, whilst a radar chart gives a snapshot summary at the end of the conversation, it could be beneficial to instead, visualize the summary in the form of Gantt chart. Fig. 26 from SR26 uses a Gantt chart metaphor to show the sequence of utterance type from both the counselor and client to facilitate the training of junior psychology counselors.

#### 5.2.2 Themes across conversations

It can be useful to compare multiple conversations according to its overall thematic content. This allows the user to quickly analyze conversations over weeks or months (Fig. 14 from SR16, Fig. 15 from SR12, Fig. 16 from SR11, Fig. 6 from SR08), or to quickly compare conversations concerning thematic content similarity (Fig. 17 from SR04). In this section, we find that the data type spans across all three categories – speech transcripts, instant messaging and emails – which suggests that the metaphors chosen were agnostic to data types.

### 5.3 Argument Structures

Visualizing the structure of arguments is helpful in clarifying the premise of an argument, and helping students learn to present their arguments better. In this subsection, all three included studies (Fig. 18 from SR10, Fig. 19 from SR02, Fig. 20 from SR23) use the graph metaphor similarly to represent the entire debate. The nodes represent arguments made, whilst the directed edges represent the "supporting" or "attacking" relationships between the nodes.

### 5.4 Emotions

Visualizing the emotional content of the conversation typically fall into two groups. Firstly, it allows the user to see an overview of whether the conversation's emotional profile is positive or negative (Fig. 21 from SR07, Fig. 22 from SR09). Secondly, it attempts to re-inject the non-verbal emotional content that is mostly absent from online conversations (Fig. 23 from SR01, Fig. 24 from SR20, Fig. 25 from SR03). There are two central metaphors used to visualize emotional content – the use of colors or avatars – each of them is discussed in turn.

The use of colors to represent emotions is intuitive and straightforward. Cimbalo et el. [57] have shown that different emotions are associated with different colors. In the following two studies (SR07, SR09), colors have been used to denote the emotions of the author.

Apart from using colors, the use of avatars to represent emotions reinjects the natural, non-verbal expressiveness of the face back into the conversation. All three studies included in this review automatically extract the emotions behind the text and render an appropriate avatar to enrich the user experience. They differ in the complexity of the rendered avatar – ranging from a line drawing of a face to an expressive 3-Dimensional animated avatar with a wide range of emotions.

## 6 Conclusion

This survey has reviewed the state of the art on visualizations of dyadic conversations. We observe that the visualizations come from a wide spectrum of domains such as health-care, customer service, and personal use. Because of this, we also observe that the lens of analysis is from a wide range from dialogue-acts to emotional content and arguments. This research area is highly interdisciplinary in nature, including conversation analysis from sociolinguistics, visualizations from computer science, understanding emotions from psychology and bespoke themes in conversations such as medicine and crisis counseling.

In the next closing paragraphs, we identify research gaps that emerged during the review. There are two main categories of gaps – (1) experiment design and (2) new visualization modules. We would discuss each of the following in turn.

### 6.1 Suggestions for future research experiment design

In the purposes section, we found that most of the studies (18 out of 25) did not design a quantitative user study to test any hypothesis. Whilst Chen & Czerwinski [58] reported that there was a rapidly growing interest in the empirical evaluation of information visualization, nearly two decades later, we observe that the visualization community has evolved to accept more than empirical approaches. Whilst quantitative approaches have its strengths, such as the ability to detect statistically significant effect sizes, such approaches are currently being debated over its applicability in human-computer interaction fields [59]. For questionnaire design, we recommend that researchers consider employing standardized questionnaire, like the SUS [49].

For those considering a qualitative approach, we recommend a separate extensive survey where Lam et al. [47] discussed the qualitative study of people's subjective feedback and opinions at length.

### 6.2 Suggestions for future visualization modules

#### 6.2.1 Lack of multi-modal visualizations with text

Most systems included in this review process the text data independently, i.e., the multi-modal analysis is mostly absent. For example, although prosody is known to influence the interpretation of the words and reveal the emotional state of the speaker, it is not used in any of the visualizations that analyze the content of the conversation. Previous research like Yang [60] visualize prosody but are not covered in this review because the visualizations do not take into account the conversation content. This lack of multi-modal visualization on top of text analysis highlights a gap in the current state of visualization research.

Of our included studies, the only exception is SR05, where an aspect of non-verbal behavior in the form of keystrokes is visualized. The lack of high-quality multi-modal datasets and the lack of mature multi-modal semantic content extraction algorithm to extract multi-modal features from visual and audio data could be plausible explanations for the absence of such visualizations.

#### 6.2.2 Visualizations for the expert users

Our analysis in section 3.1 revealed that there are two potential gaps in the visualization research of dyadic conversations. Firstly, there is no analysis of the emotional content aimed at expert users in the health care domain. There could be research opportunities here as addressing patient emotion is an essential aspect of conversations in the health care setting [61], [62]. There is some existing research that has applied affective text analysis onto doctor-patient communication to identify features that are predictive of excellent doctor communication ability [63] however, such models are not given as feedback to doctors in the form of visualizations to suggest possible areas of improvement.

Secondly, existing visualization research only targets experts from the health-care domain. We posit that other professional dyadic conversations could benefit from visualizations, examples include a professional consultation with a lawyer or an earnings interview between the analyst and the senior manager of the company.

#### 6.2.3 Text analysis methods

Disclosure of the underlying extraction mechanism of theme, argument structure, and emotional content is often limited. The technical details of the extraction are not described in detail for many studies, of those who do disclose, we observe that the techniques range from simple count heuristics to more advanced models like Latent Dirichlet Allocation [64], Hidden Markov Models, and neural networks.

For emotion detection systems, the current systems are limited to detecting the emotions behind each utterance which might be too localized to be helpful. We suggest that it would be beneficial for systems to go broader and provide a visualization for the overall mood of the person.

In conclusion, as digital conversations continue to scale, continual, thoughtful and practical explorations in this space will help conversation participants, as well as third-party analysts, grapple the analysis of the unprecedented amount of digital conversation data.

**Joshua Y. Kim** is a Ph.D. candidate in the Faculty of Engineering and Information Technologies of the University of Sydney. He is a data scientist with an interest in applying machine learning techniques to better understand conversations.

**Rafael A. Calvo** is Professor at the University of Sydney, ARC Future Fellow and Director of the Software Engineering Group that focuses on the design of systems that support wellbeing in areas of mental health, medicine and education. He has a PhD in Artificial Intelligence applied to automatic document classification and has also worked at Carnegie Mellon University, Universidad Nacional de Rosario, and as a consultant for projects worldwide.

**Kalina Yacef** is Associate Professor at the University of Sydney; her research lies in the fields of Artificial Intelligence in Education, Educational Data Mining and Human Computer Interaction. Her research aims to create smart, personalised computing systems to support learning and teaching, with a particular focus on creating ways to mine the rich stream of interaction data between learners and computer systems and build interfaces to control this data and visualise results.

**N.J. Enfield** is Professor of Linguistics at the University of Sydney and director of the Sydney Social Sciences and Humanities Advanced Research Centre, and the Sydney Centre for Language Research. He is head of a Research Excellence Initiative on The Crisis of Post-Truth Discourse. His research on language, culture, cognition and social life is based on long term field work in mainland Southeast Asia, especially Laos.

# APPENDIX

## Appendix 1 – Copyright permission

The following table provides details about permissions to reproduce the screenshot of the visualization in this review. We acknowledge and thank the authors for the permissions granted to make this review possible.

Table 5
Copyright Permissions Summary

| Included study | Rightslink License #<br>Source of permission |
|---|---|
| SR01 | 4406791451151 |
| SR02 | 4313870488146 |
| SR03 | Open access |
| SR04 | 4313891362752 |
| SR05 | Author's email |
| SR06 | 4406790233292 |
| SR07 | 4406790470990 |
| SR08 | 4406790765288 |
| SR09 | Open access |
| SR10 | 4406790949374 |
| SR11 | Open access |
| SR12 | 4406791098778 |
| SR13 | 4314151188783 |
| SR14 | Open access |
| SR15 | Open access |
| SR16 | Open access |
| SR17 | Open access |
| SR18 | 4406791237464 |
| SR19 | 4313900586774 |
| SR20 | 4313860949075 |
| SR21 | Not applicable |
| SR22 | Not applicable |
| SR23 | 4313870022205 |
| SR24 | Open access |
| SR25 | Open access |
| SR26 | 4493500619842 |

## Appendix 2 – Screenshots of visualizations

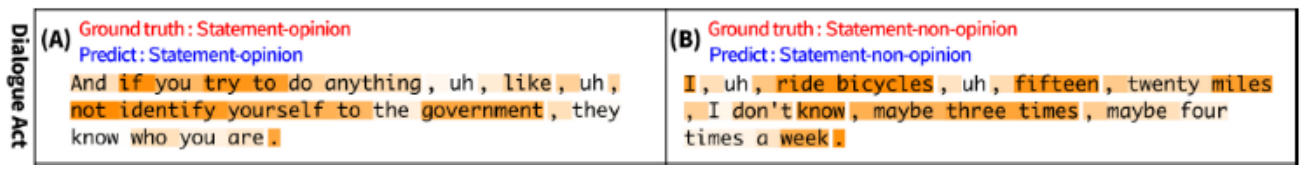

Fig. 2 – Visualization for dialogue act classification task from SR17. The darker color a word gets, the higher attention weight it has contributing towards the prediction. The texts in red represent the ground truth (i.e., the correct category), and blue represents prediction.

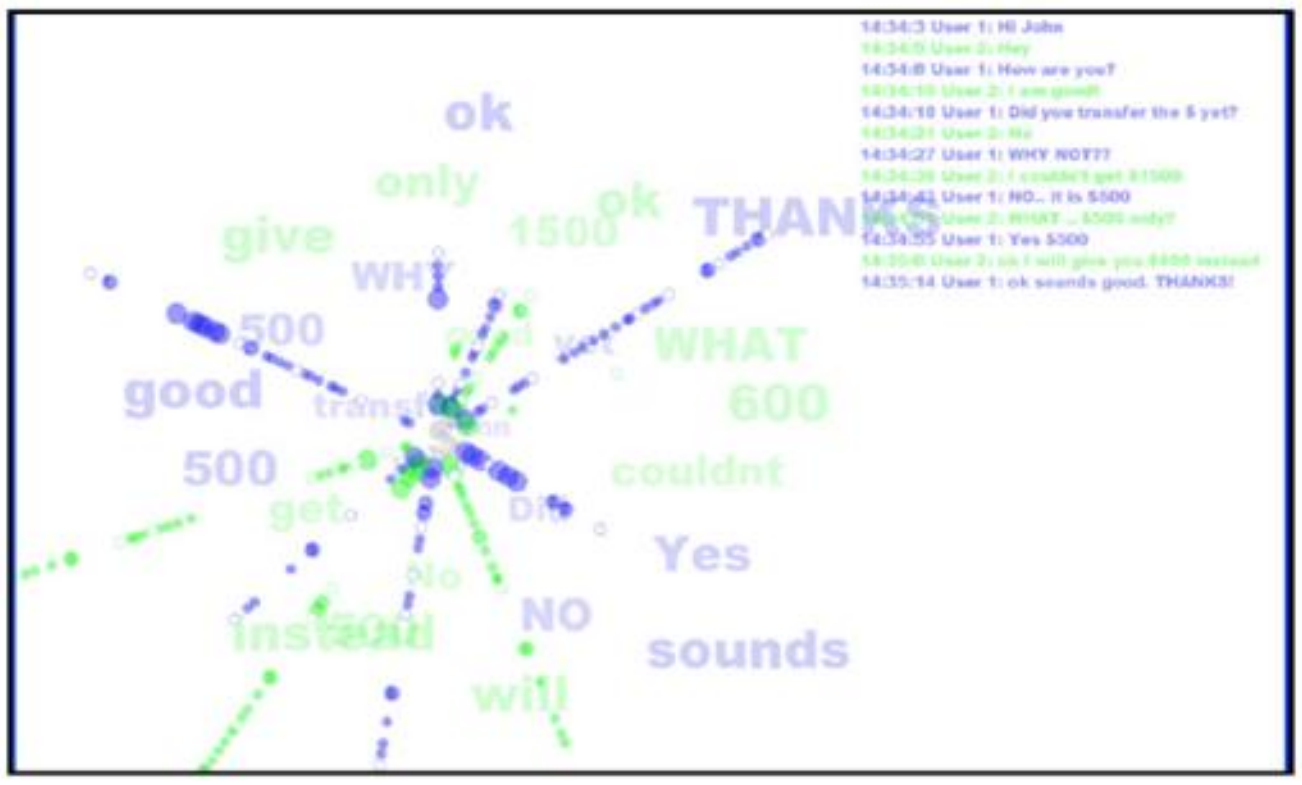

Fig. 3 – ChatViz sample live-chat visualization from SR05. Color represents speaker, and color intensity represent age in conversation. Recent words appear larger. The line of bubbles indicates keystrokes. The single-character symbol (‘$’) represent the topic of the conversation - ‘$’: money oriented, ‘?’ question oriented, ‘!’ exclamation intensive.

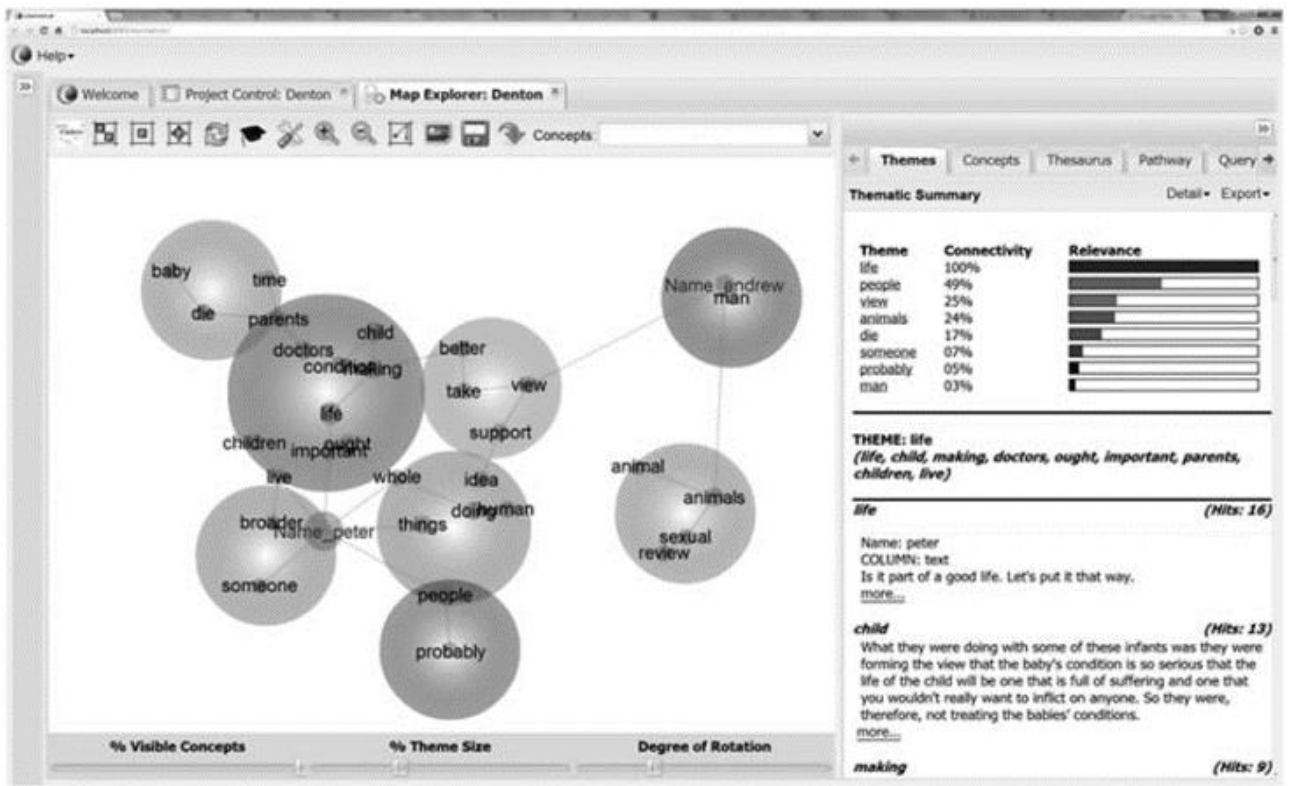

Fig. 4 – In the Leximancer system of SR13, the visualization illustrates the connections between words using a graph metaphor – with the node representing conceptual words and edges representing the conceptual similarity between words. The interactive visualization also shows how pairs of concepts are used in the original text.

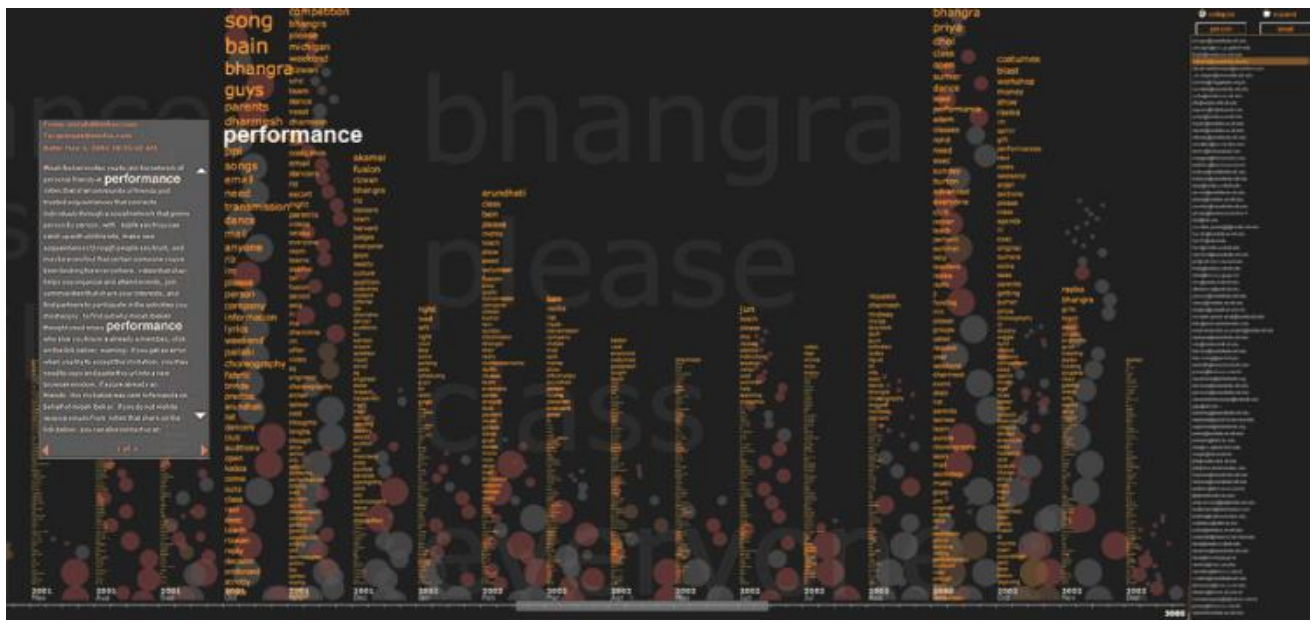

Fig. 5 – In Themail of SR24, terms used over a year are visualized as large feint words in the background. Monthly words appear as yellow words in monthly columns in the foreground. The words are selected and sized according to the TF-IDF algorithm. For the bubbles, the size indicates the length of the message whilst the color indicates whether the email is incoming or outgoing.

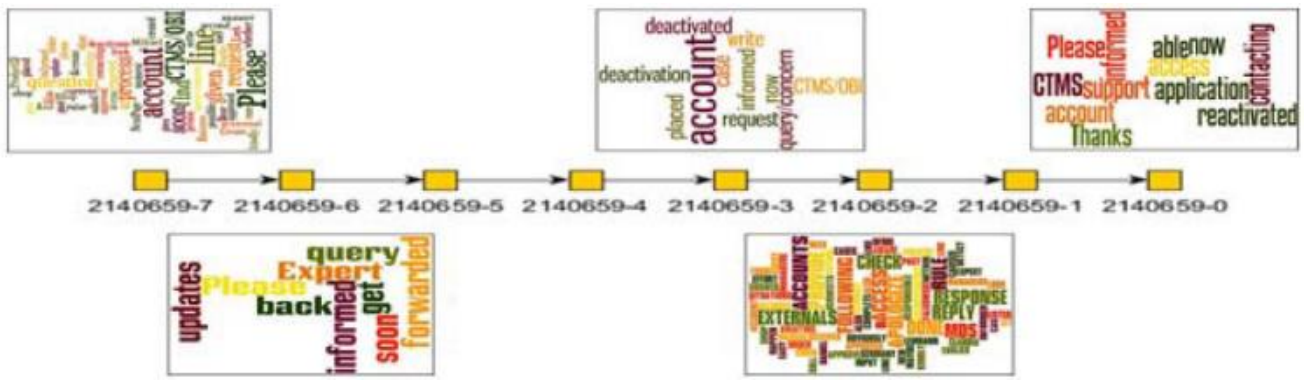

Fig. 6 – SR08 utilizing word clouds to summarize each email in a complaint conversation. The size, orientation, and color representation are not disclosed.

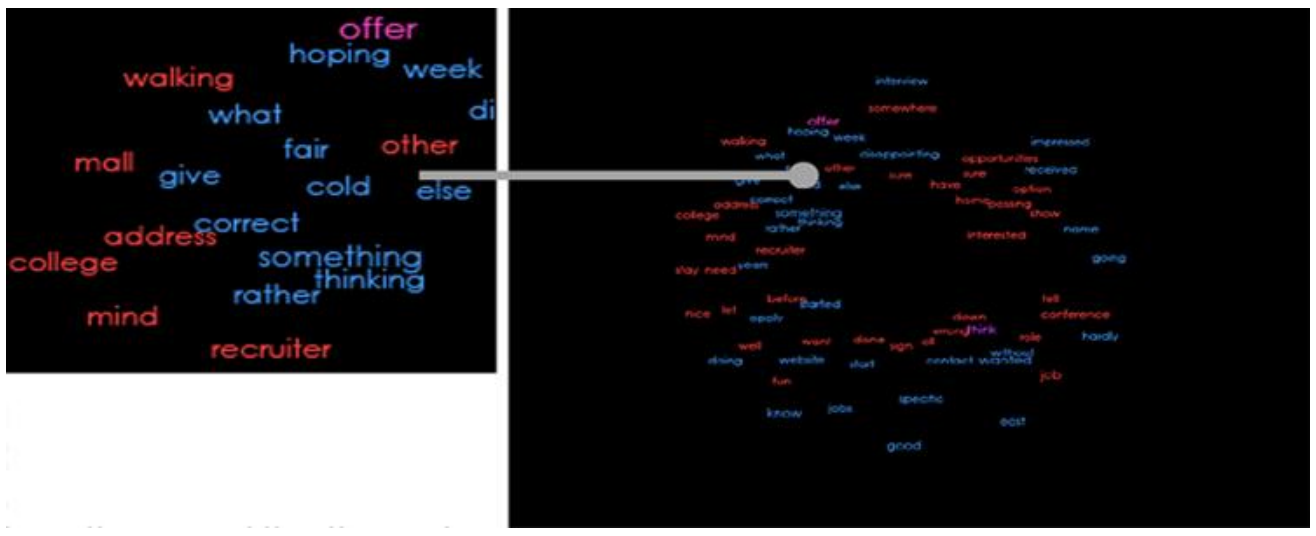

Fig. 7 – SR25 visualizes voice conversations in real-time. The color indicates the speaker, the pink/purple represents both speakers have used these words. Every minute, the words move outwards, with newer utterances on the inner circles.

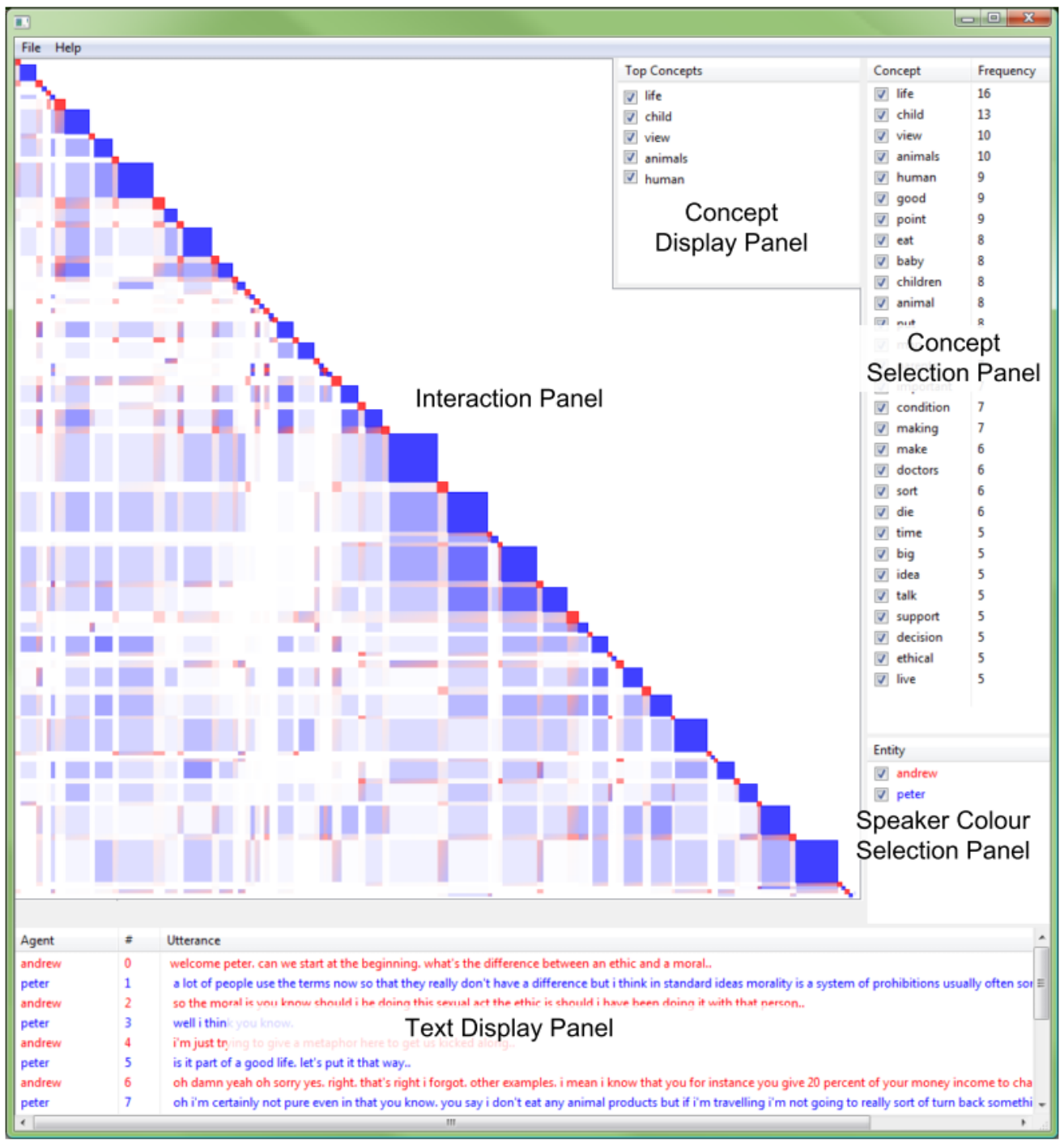

Fig. 8 – Conceptual recurrence plot from Discursis, SR06, visualizing a television interview transcript. In the main window, one box on the diagonal represents one utterance, color-coded by the respective speaker, and its size represents the length of the utterance. Off the diagonal, the boxes represent a common subject matter between two utterances – the utterance that is directly above itself and directly to the right of itself – the color coding still represents the speaker, with a mix of colors used when it applies to both speakers. In the side panels, the user could filter for a concept, as illustrated in the next figure.

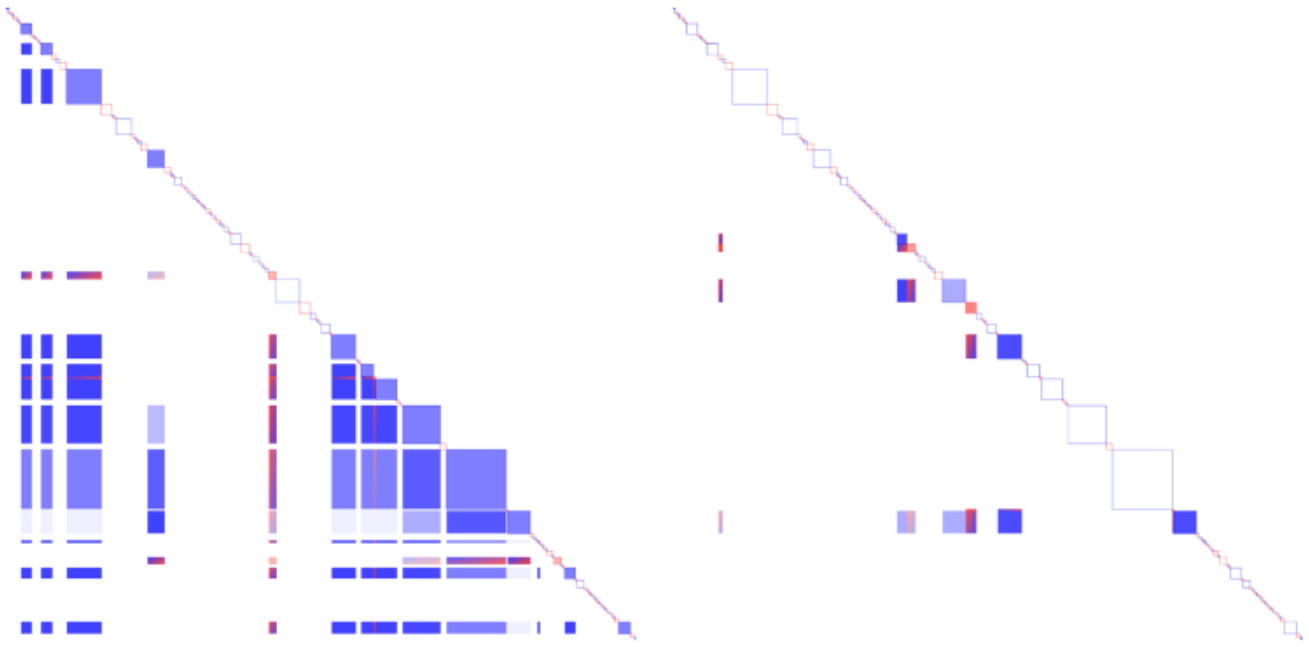

(a) Depression (b) Politics

Fig. 9 – Another example plot from Discursis, SR06, visualizing a television interview. (a): only filtering for depression related recurrence and (b) only filtering for politics-related recurrence. This comparison shows that the blue speaker wishes to talk about depression at length (notice the size of the diagonal boxes are big); the red speaker tries to initiate questions about politics, but there is little participation from the blue speaker.

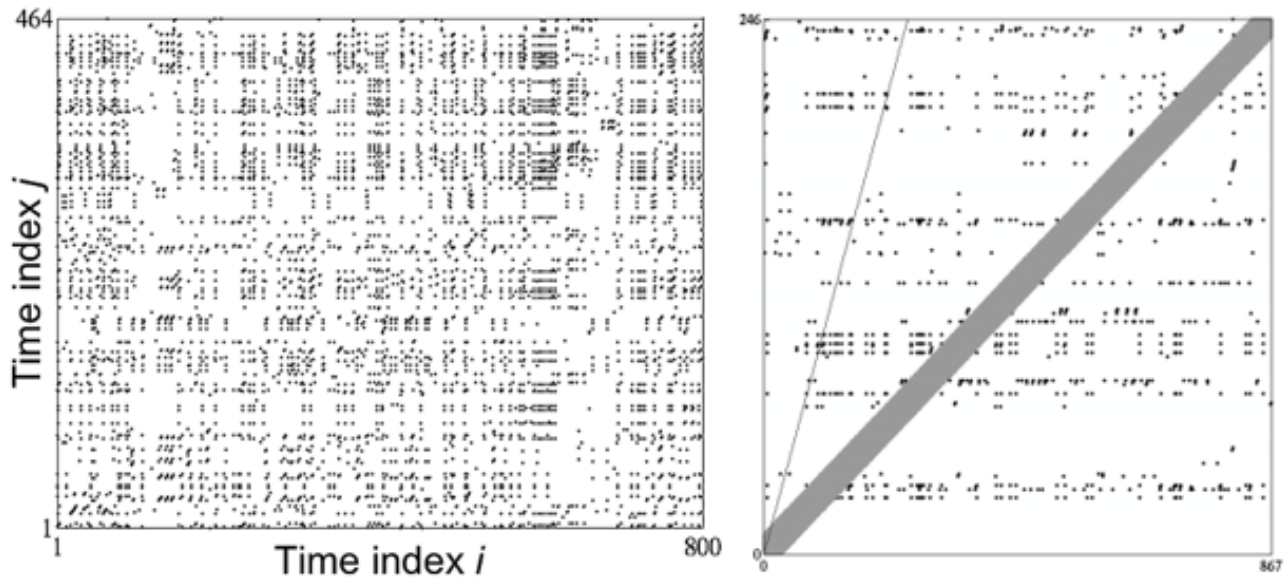

Fig. 10 – Recurrence plot from SR19. (Left) A filled pixel indicates where there are matching syntactic bigrams occurring. Syntactic bigrams are combinations of two syntactic components, for example, the syntactic bigram of the phrase 'beautiful car' is 'adjective-noun'. In this study, the authors were investigating whether a child (represented on the y-axis) is learning syntactic bigrams from the caregiver (represented on the x-axis), and thus designed this visualization. For example, the caregiver used the phrase 'beautiful car' in turn 1 and the child used the phrase 'delicious apple' in turn 10 since both phrases are of the syntactic bigram 'adjective-noun', the pixel in position (1,10) would be filled. Therefore, the density of filled pixels gives an indication of the syntactic content similarity between the two speakers. (Right) An illustration of another conversation with the diagonal line of incidence and its band-size plotted.

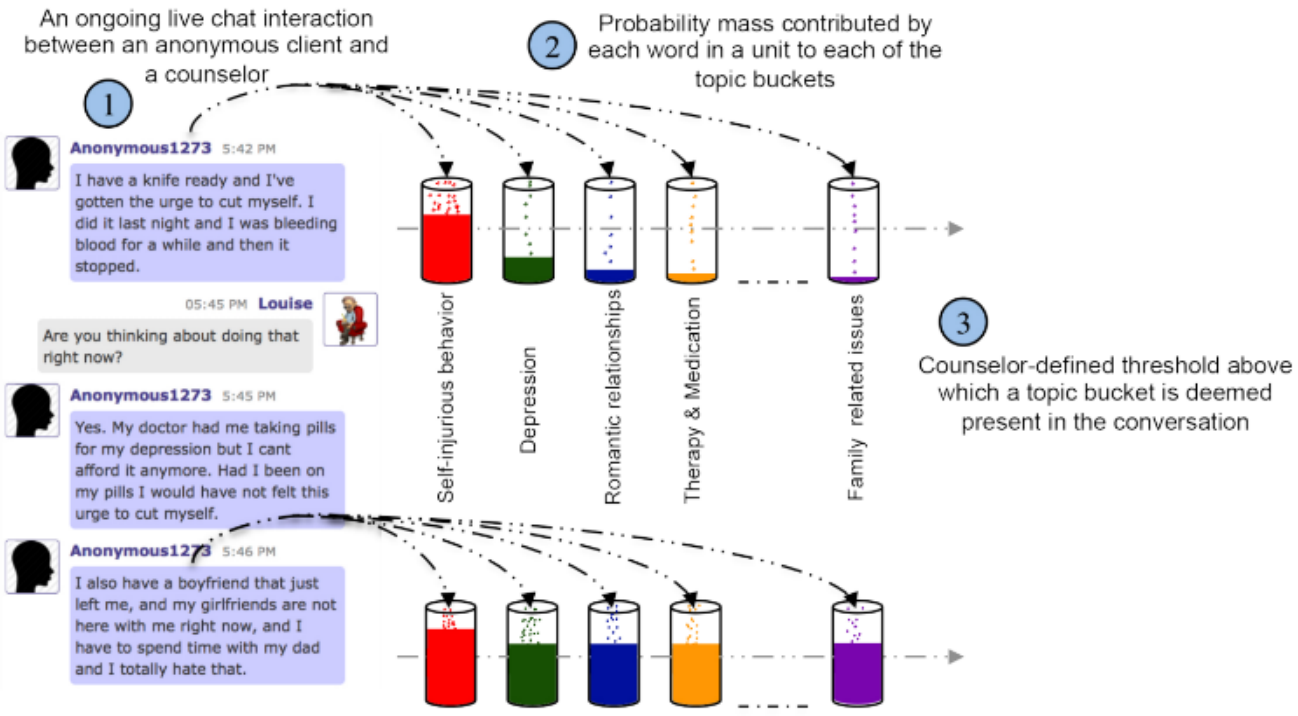

Fig. 11 – Visualization from Fathom, of SR14. The crisis counselor participates in a text messaging conversation with the caller. Words in the text messages each contribute towards the topic-buckets. The topics are first agreed with a group of prevention science psychologists before the model is trained. The counselor can also define the threshold above which a topic is deemed to be present. The screenshot shows that the first message adds more mass to self-injurious behavior than the third message, plausibly because the words "cut myself" is used in the first message.

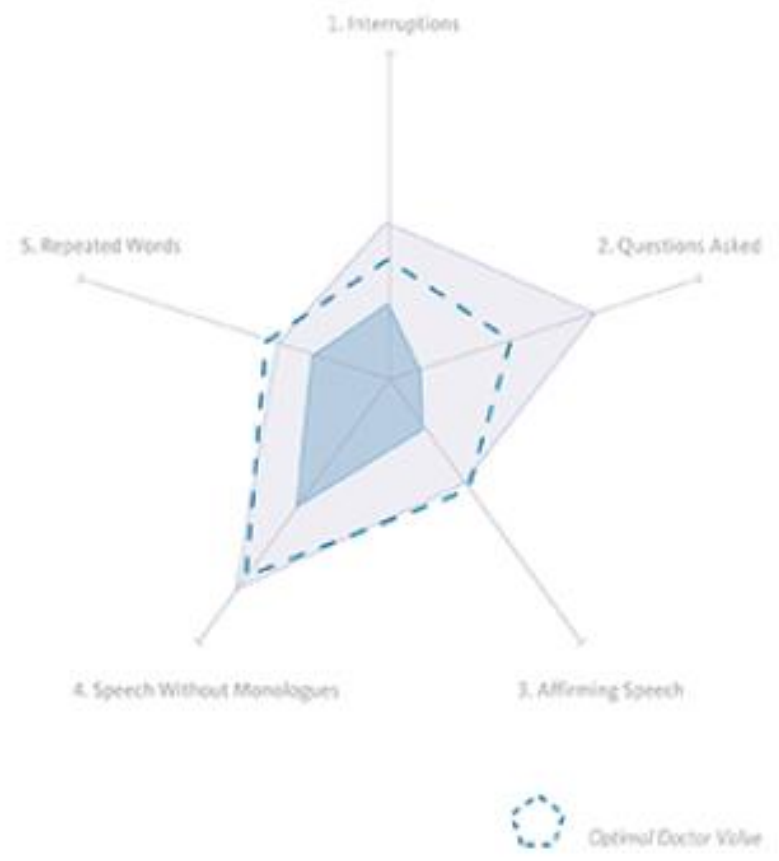

Fig. 12 – A radar chart visualization from Monologger of SR15, visualizing a doctor-patient consultation. The five axes are (1) Interruptions, (2) Questions Asked, (3) Affirming Speech, (4) Speech Without Monologues and (5) Repeated Words. The dark (light) grey region is the performance of the doctor (patient), whilst the dotted line denotes the optimal doctor value.

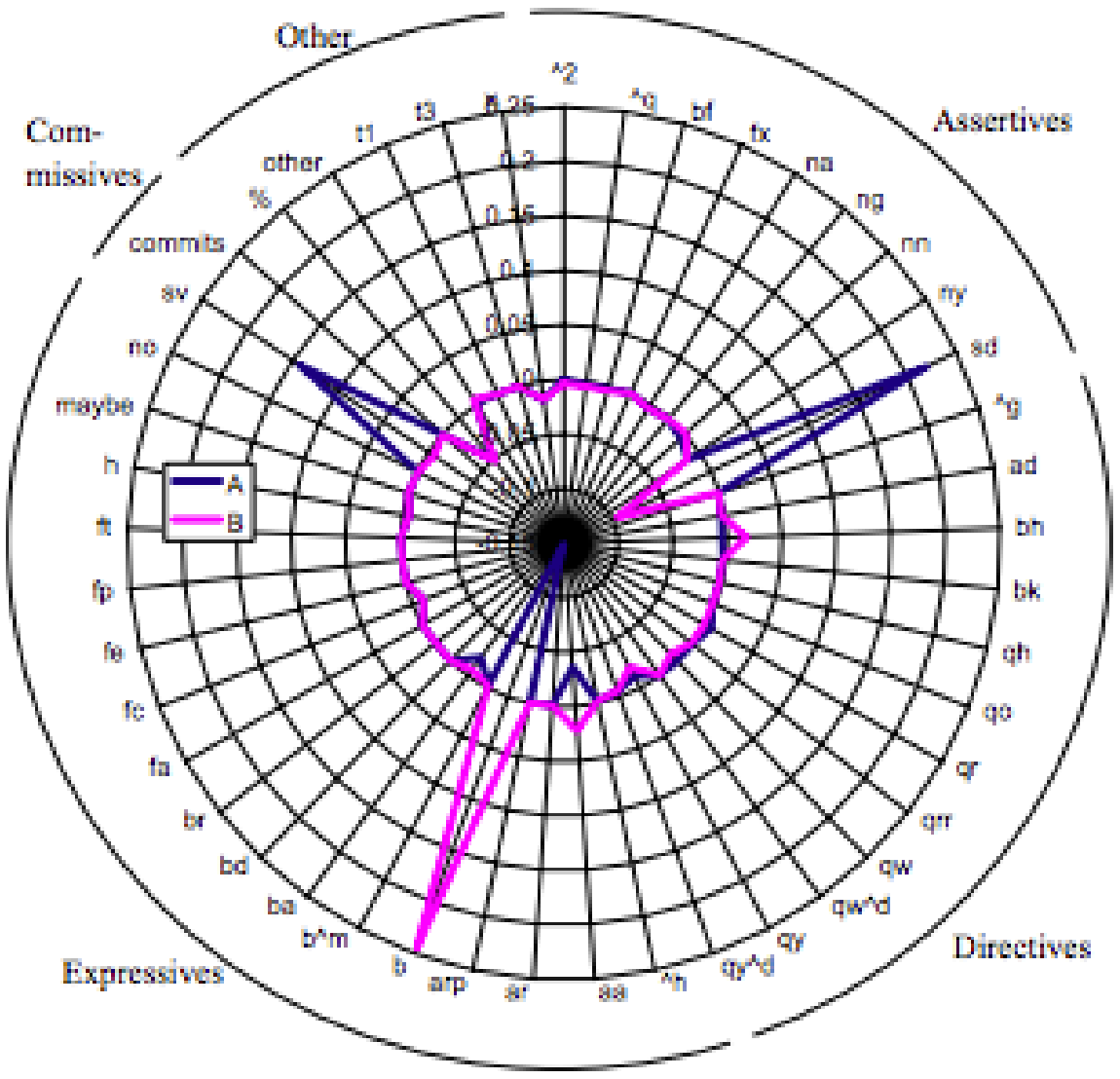

Fig. 13 – Speech act profile radar plots from SR18 for two speakers of a speech transcript. Speech acts are also sometimes known as the illocutionary act, which focuses on the force or intent of an utterance. The 42 speech acts are grouped into five categories as per Searle [65]. Here, we observe that the one speaker dominated the conversation, which is indicated by the high statement (sd) count by one speaker and low statement (sd) count plus high backchannel (b) count by the other speaker.

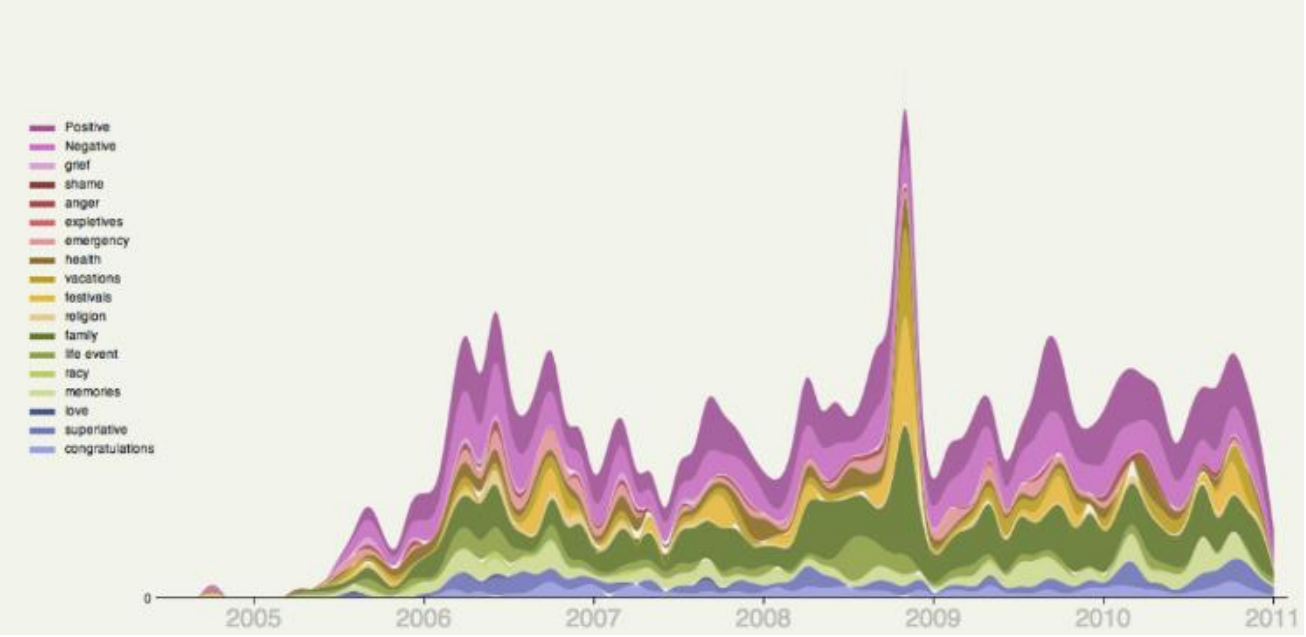

Fig. 14 – MUSE from SR16 can visualize the topics that are prevalent in emails with a specific person for certain time periods. The Y-axis here represents the frequency. To detect the topics, the authors developed a lexicon that consists of 20 topics ranging from emotions, family, health, life events for example, and use these terms to identify topics. To detect sentiment, the authors used SentiWordnet [66] and LIWC [67]. The treatment of emails having multiple topics/sentiment – e.g., an email about vacations, family, anger, and grief – is not discussed.

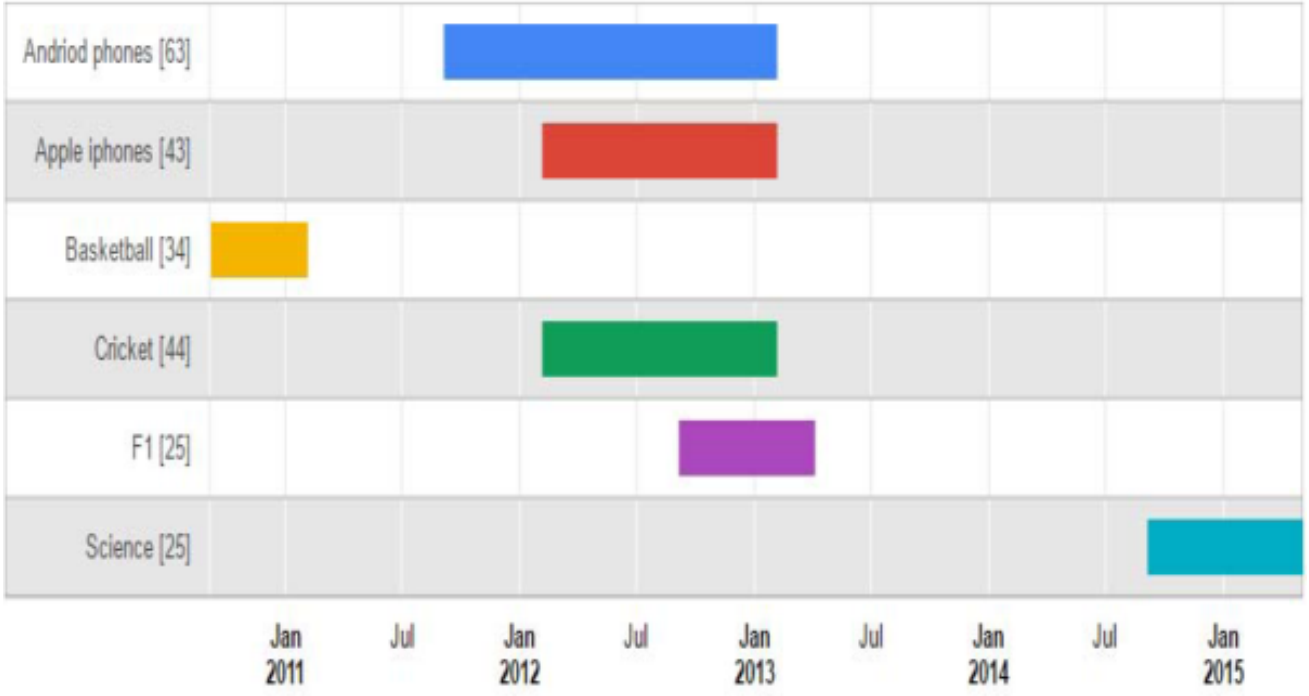

Fig. 15 – An example visualization from SR12 that visualizes the topics covered in a chat session. It uses the x-axis to represent time. However, instead of stacking the counts, the authors separated the topics using rows, and color in periods of time where a topic is discussed. This metaphor is like a Gantt-chart. The numbers in square brackets indicate the number of messages sent to all his/her contacts.

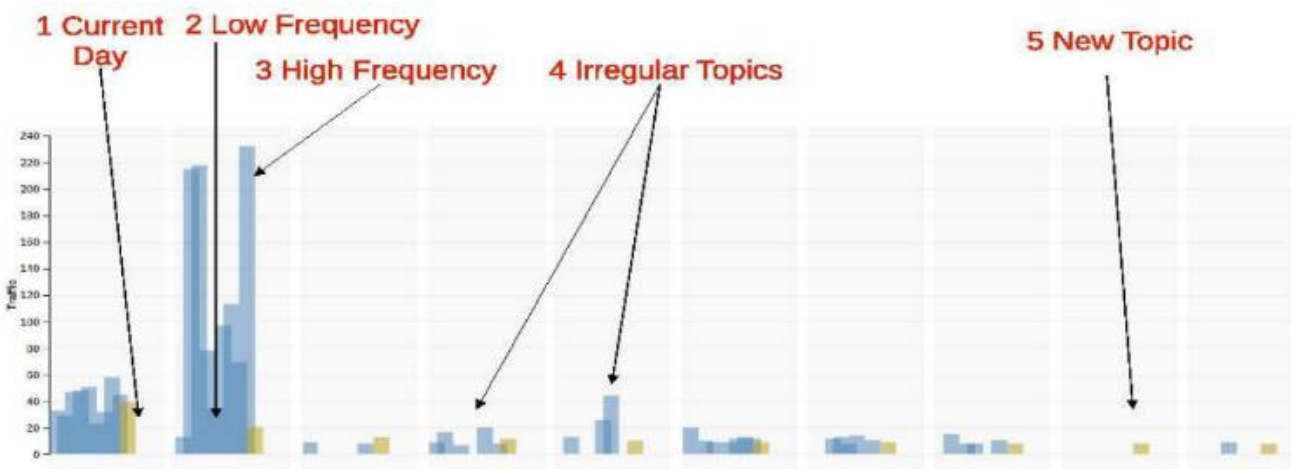

(a)

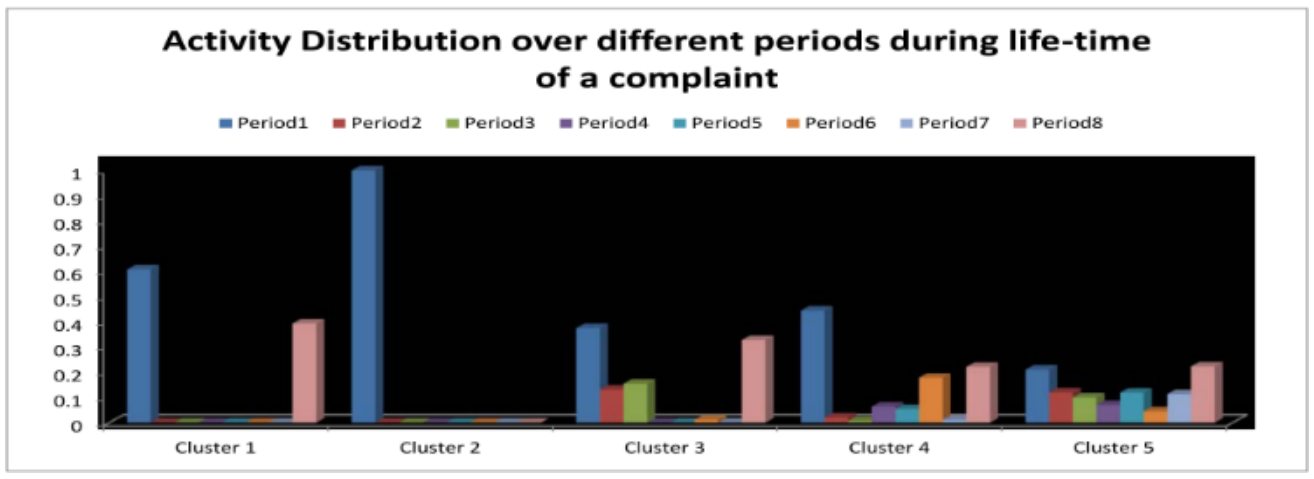

(b)

Fig. 16 – An example visualization from SR11 (a) and SR08 (b) visualizing the topics of emails. Like the previous figures, both visualizations use the x-axis to represent time and categories. (a) A group of bars that horizontally overlap with its neighbors indicate the daily count of the number of emails with one topic, with the latest day on the right. A new topic could thus be revealed by a solitary bar without any neighboring counts before. We found the visualization hard to read with the horizontal overlap. (b) is like (a), but without the horizontal overlap.

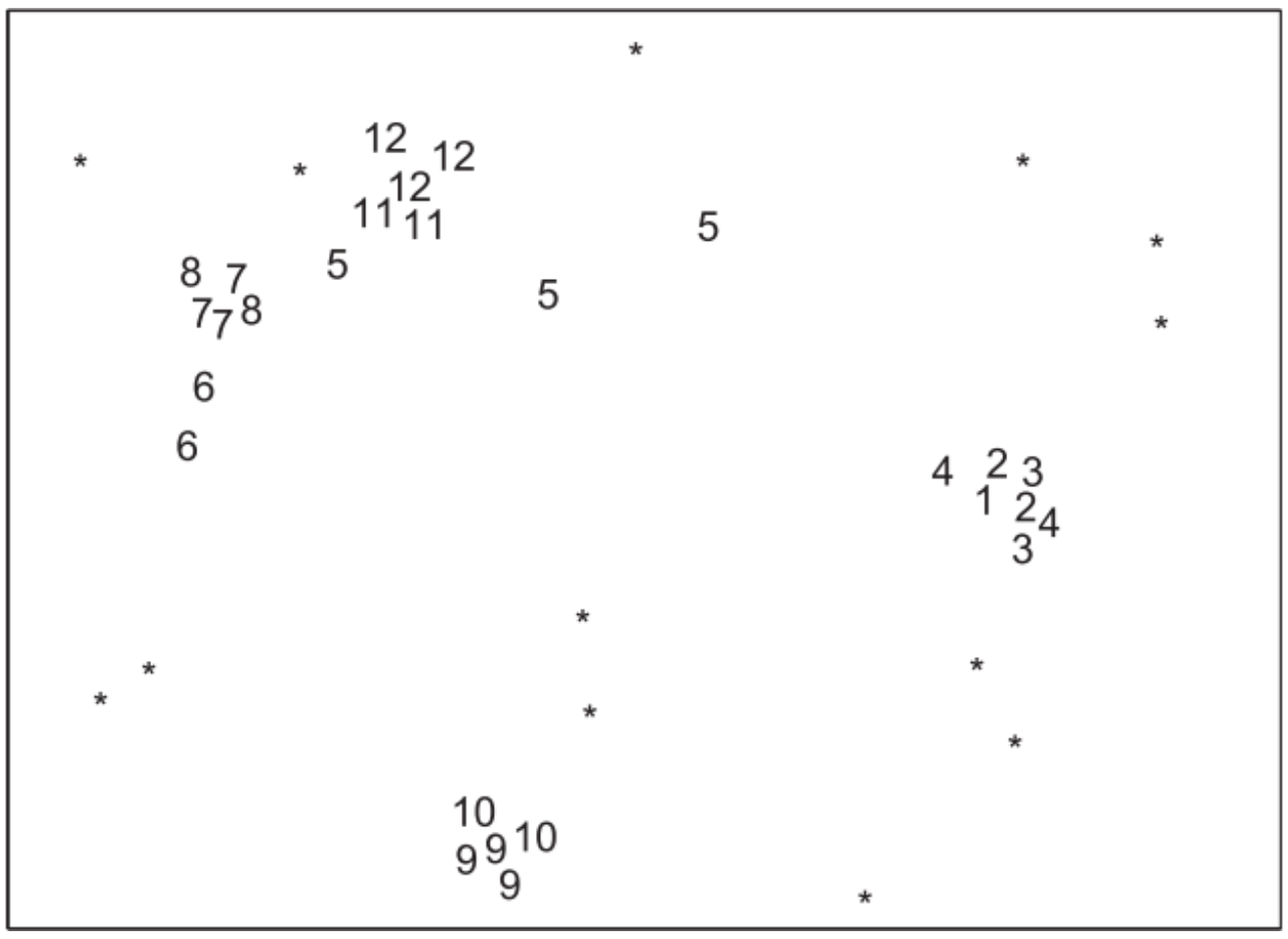

Fig. 17 – A 2-Dimensional projection representing phone conversation transcripts from SR04. A point (labeled * or with an integer) represents a conversation. The integer labels indicate which topic does the conversation belong to and the asterisk label indicates that the conversation belongs to a miscellaneous topic. Here, we observe that conversations belonging to the same topic are clustered close to one another.

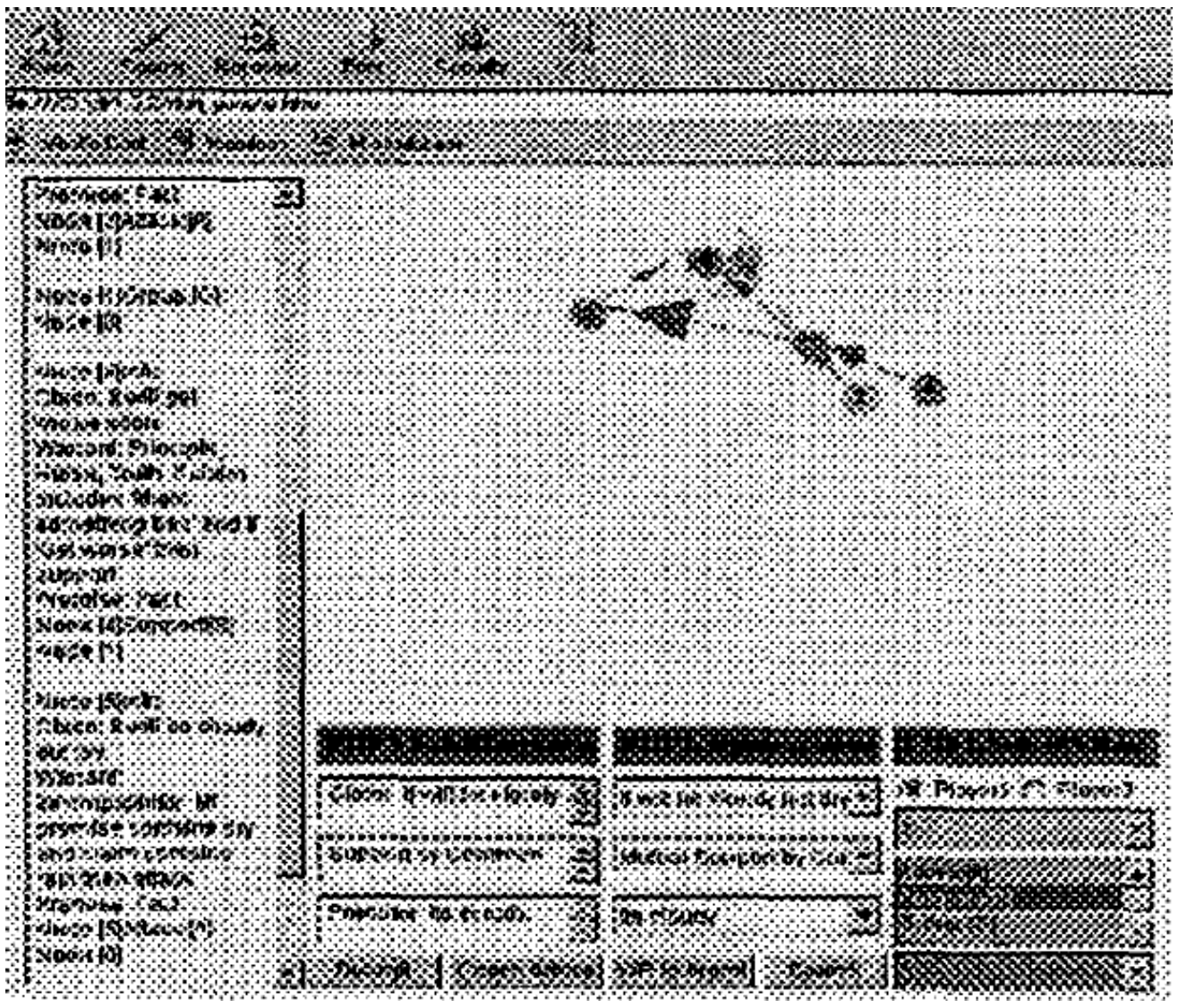

Fig. 18 – SR10 is a visualization built in 2000, the model is built on Toulmin's schema of argumentation, and includes elements like warrants, claims, premises, backing, and rebuttal. Nodes represent premises, whilst edges represent the relationship between nodes, either supporting or attacking. The strength of edges depends on the warrants. The image is of low-resolution, but we observe that the metaphor is graph-like.

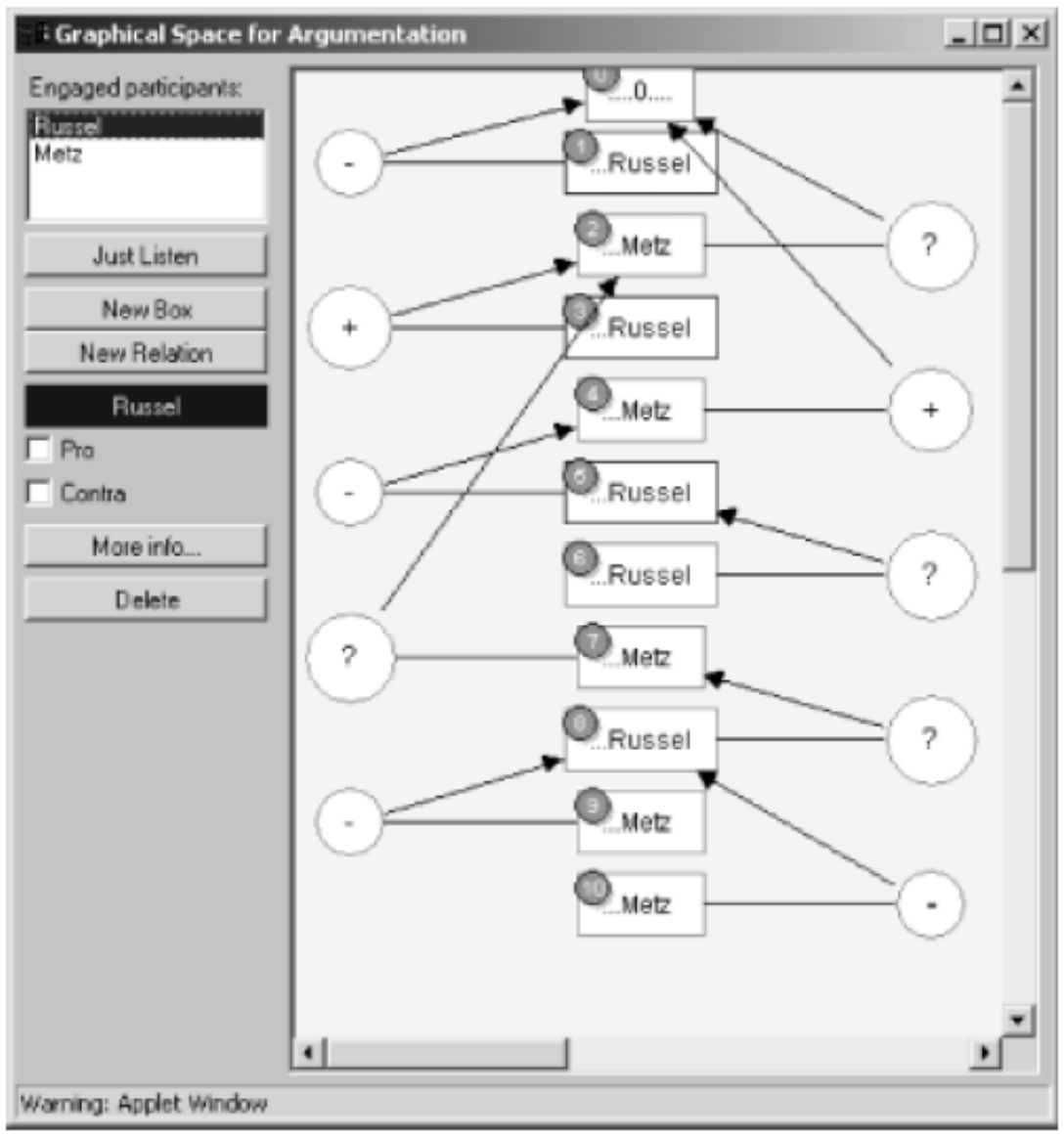

Fig. 19 – DREW visualization from SR02. For each response, the user needs to select a template to respond to the argument. For example, statement 2 does not make sense because [user's reason]. Using the template input, DREW creates a graph-like metaphor. Rectangular boxes represent the user input parts of the argument (which is shortened to only the author name in the screenshot); directional arrows make explicit the reference of the argument to the thesis. On the edges, the argumentative orientation is indicated ("+": for, "-": against, "?": undefined).

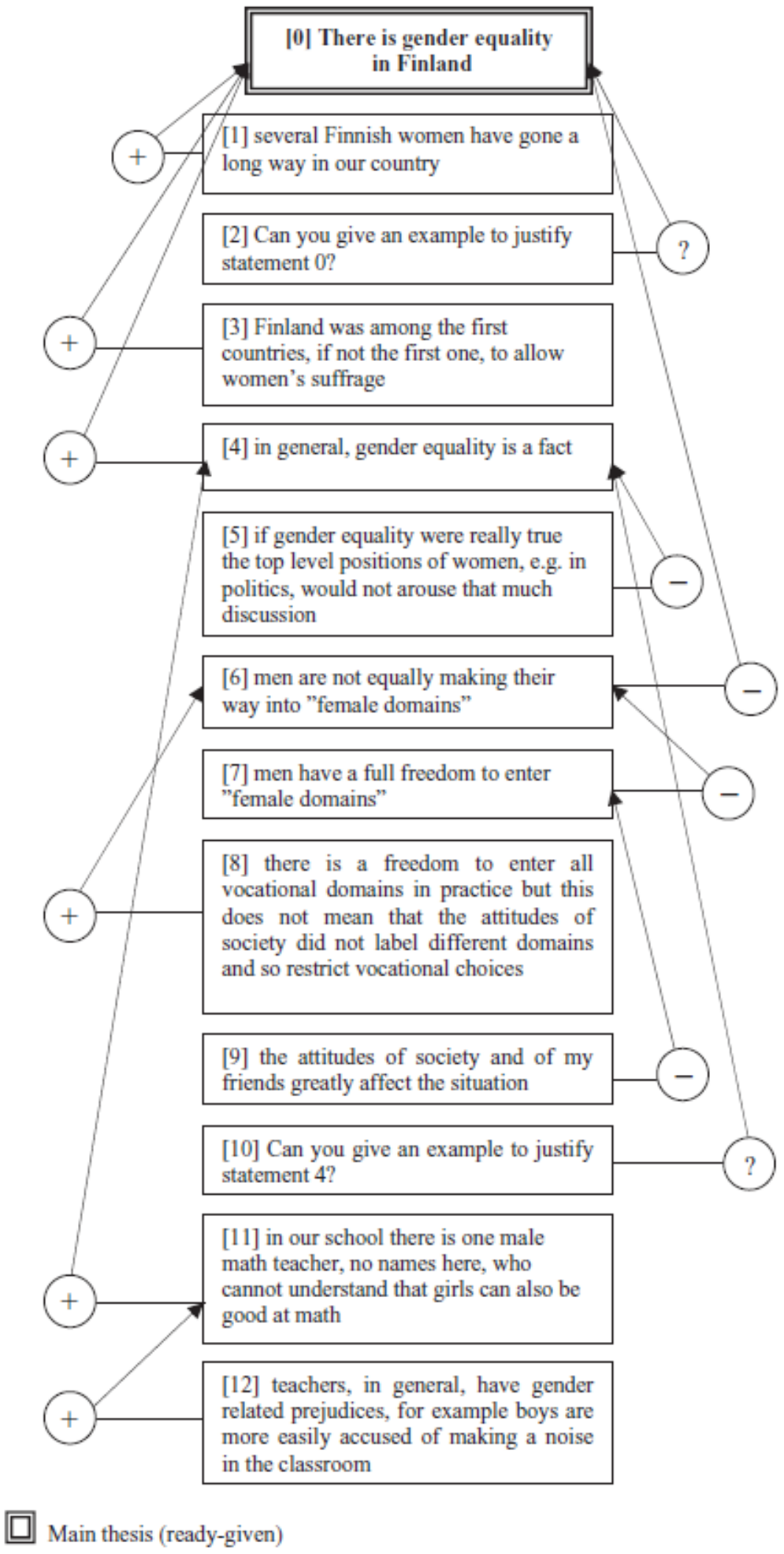

Fig. 20 – Similar to SR02, users label the relationship between arguments – supportive (+) or critical (-) – using a template. With the template, argument diagrams are automatically created in SR23.

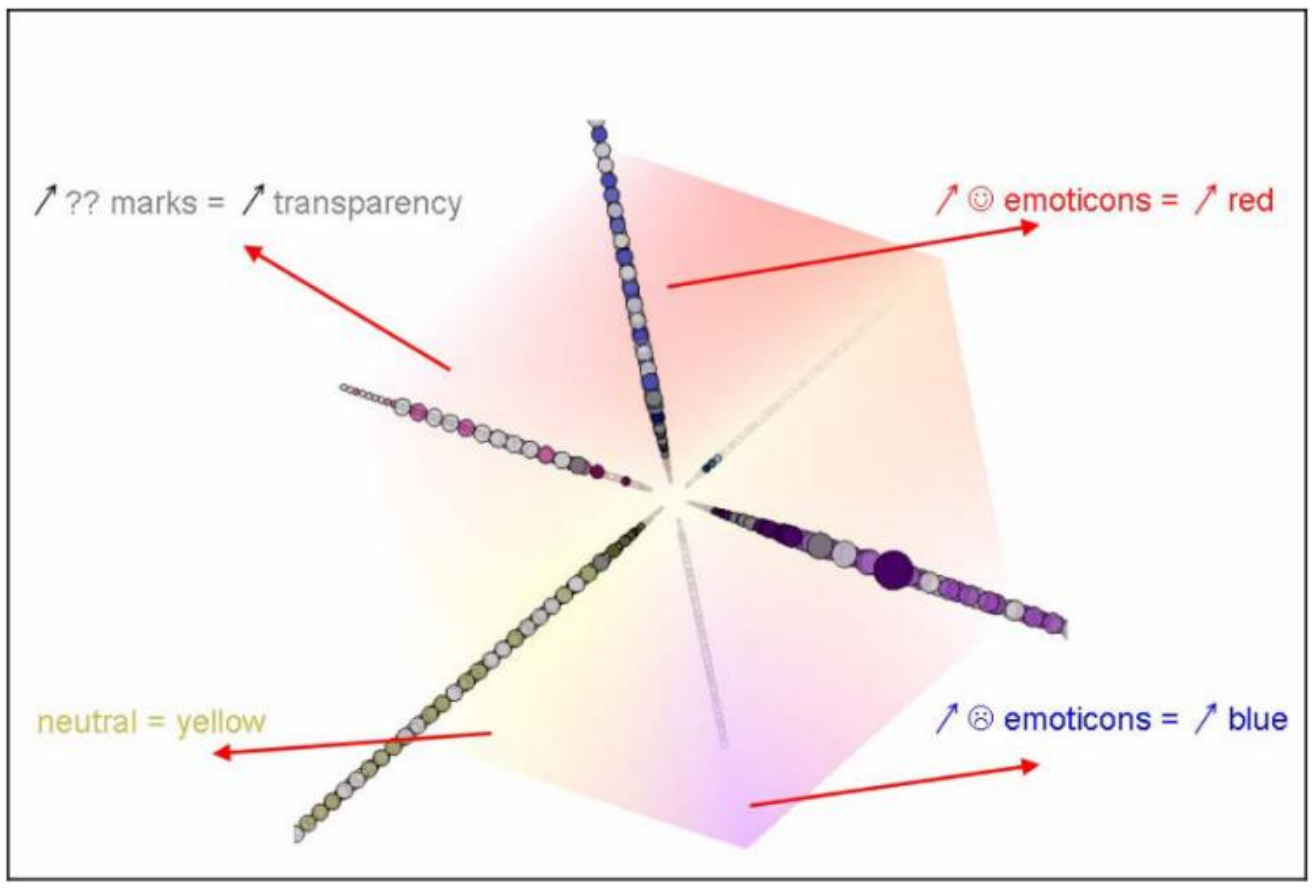

Fig. 21 – An emotion-centric visualization from CrystalChat of SR07. The visualization illustrates the emotion-content of six dyadic conversations with six different people, as indicated by the six spokes (i.e., six lines of bubbles). The bubbles are colored to identify the speakers (not emotions). The grey bubbles are spoken by the user of the visualization, whilst non-grey bubbles are spoken by the other party. In the background, there is a translucent-colored hexagon. The color of the translucent hexagon on which the line of bubbles sits upon indicates the emotional content. The color representations are as follows: Pale Yellow – Neutral; Blue – Sad (derived by counting the number of sad emoticons); Red – Happy (derived by counting the number of happy emoticons); Transparency – Ambiguity (derived by counting the number of '?' characters). For example, the conversation with the pink user is on a transparent color of the hexagon, which means that the conversation contains a relatively high level of ambiguity.

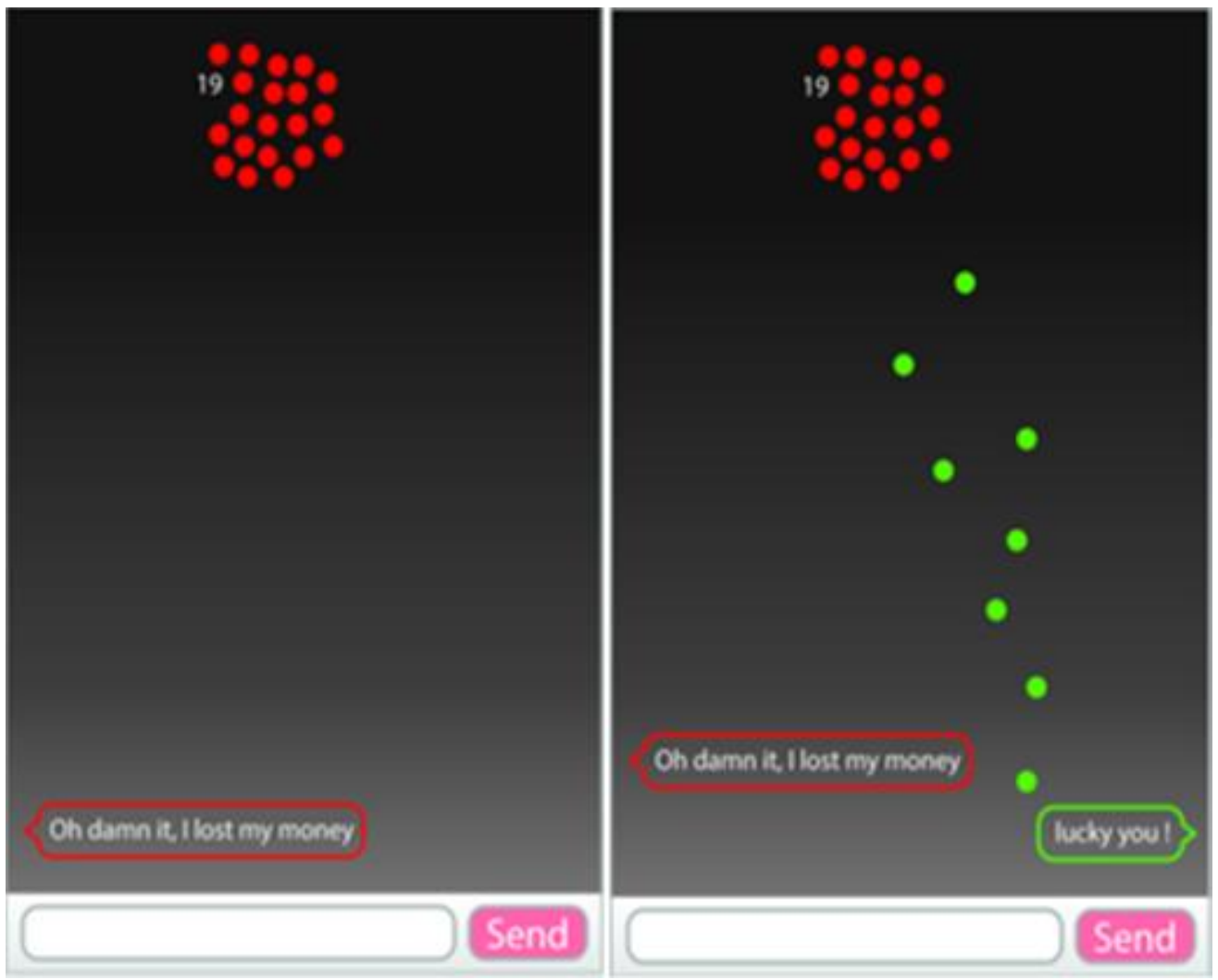

Fig. 22 – GamIM from SR09 colors the bubbles according to its positive (in green) or negative (in red) content. The content is extracted via a Naive Bayes classifier trained on the NPS Internet Chatroom Conversations Corpus of Linguistics Data Consortium. The number of bubbles is proportional to the length of the message. The top display area collects all bubbles from previous messages, and newly generated bubbles from a message would rise into the top display area – a design inspired by the lava lamp.

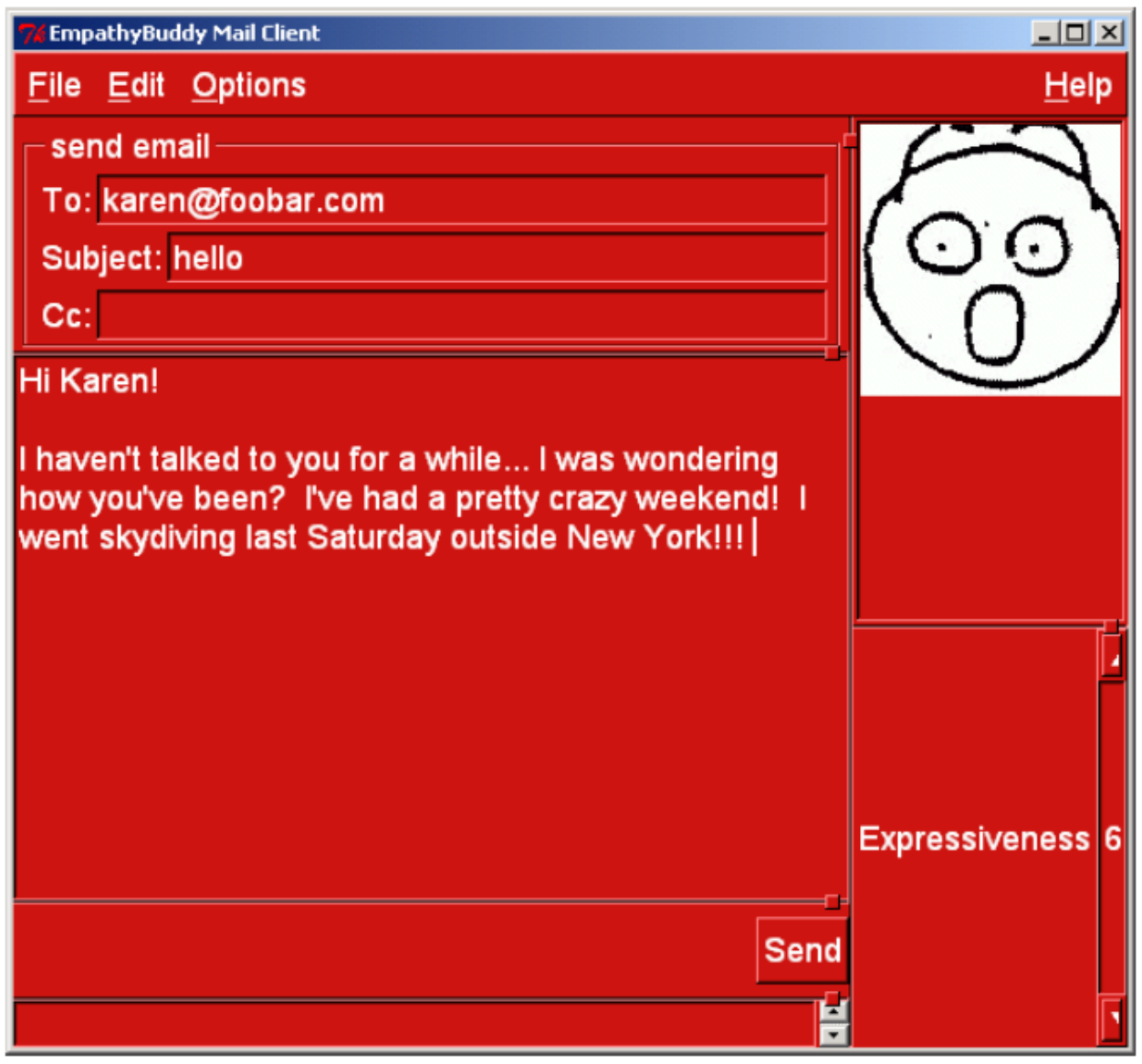

Fig. 23 – EmpathyBuddy from SR01 renders a line drawing based on the emotions detected in the email. The system can detect a total of six different emotions – happy, sad, anger, fear, disgust and surprise.

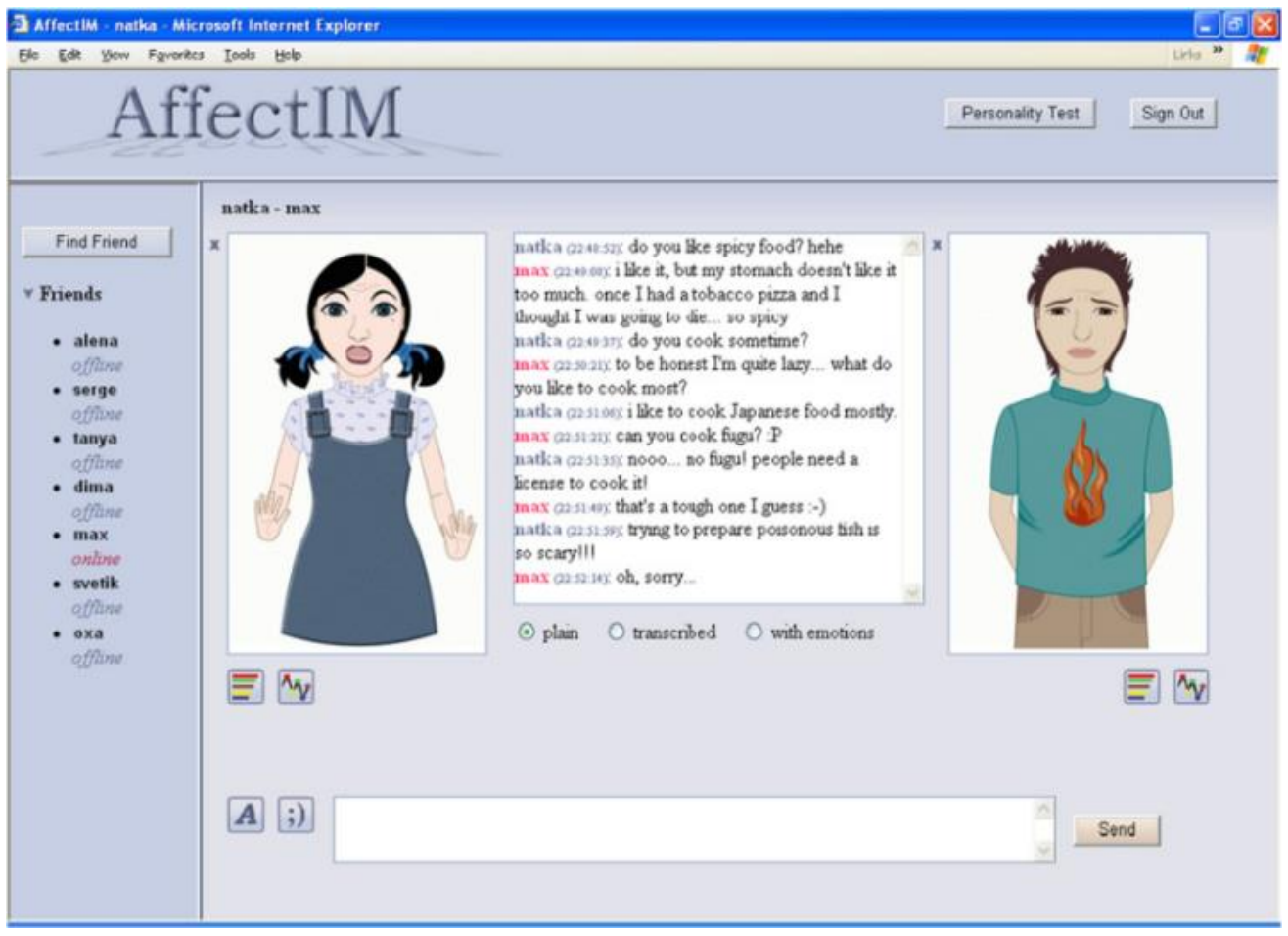

Fig. 24 – AffectIM of SR20 matches the avatar to the emotions from utterances of an instant messaging platform. A total of nine emotions could be detected – anger, disgust, fear, guilt, interest, joy, sadness, shame and surprise. After emotions detection, a separate module tempers the expressiveness of the avatar through the extraversion score, which could be set manually by the user or deduced via a short extraversion test. There is a male and female version of the avatars.

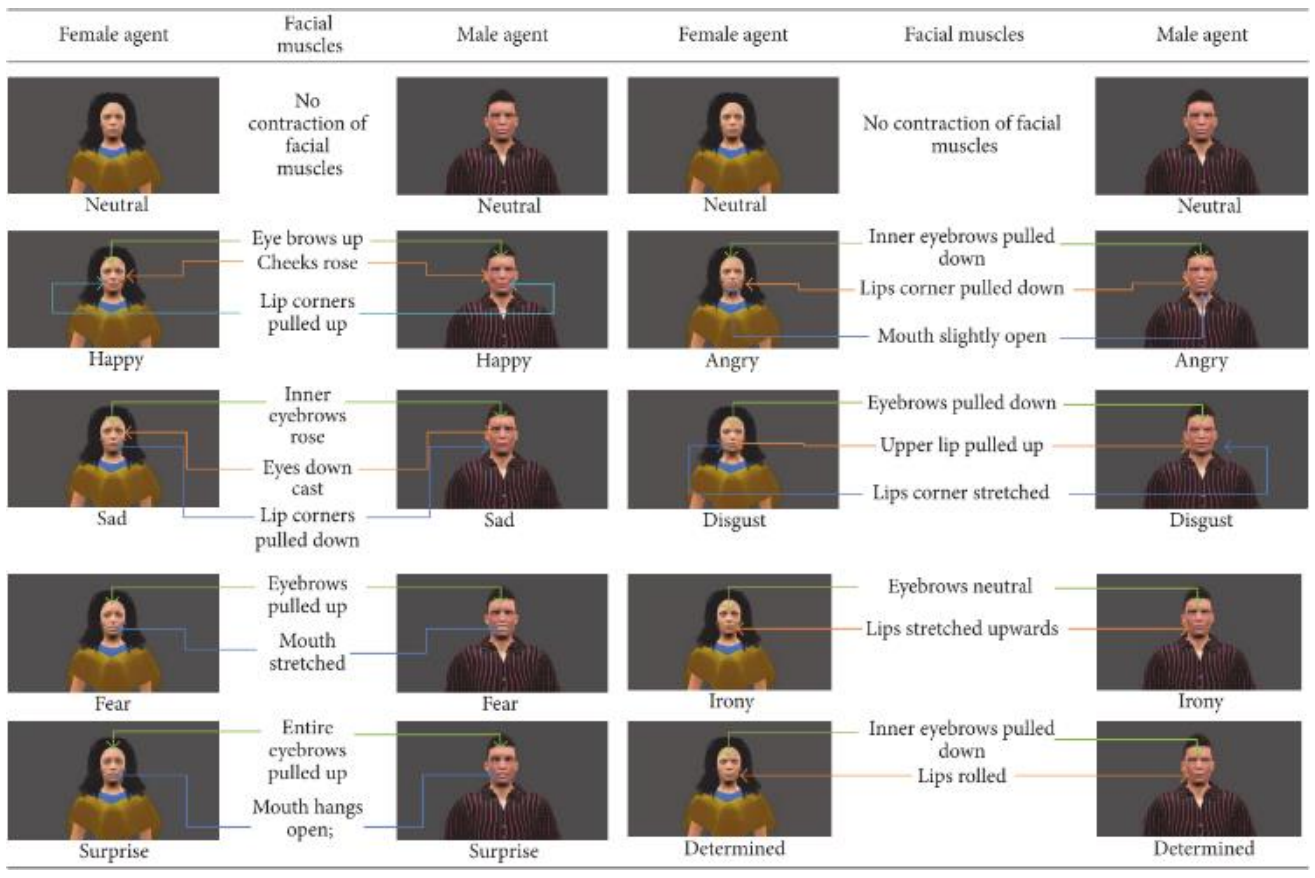

Fig. 25 – 3D avatars generated in SR03. The avatar reflects the latest utterance in the instant-messaging platform and has a male and female version of the avatar. The system could detect seven emotions – happy, sad, angry, fear, surprise, irony and determined. The avatars also have an element of naturalness injected via eye-blinking and head movements.

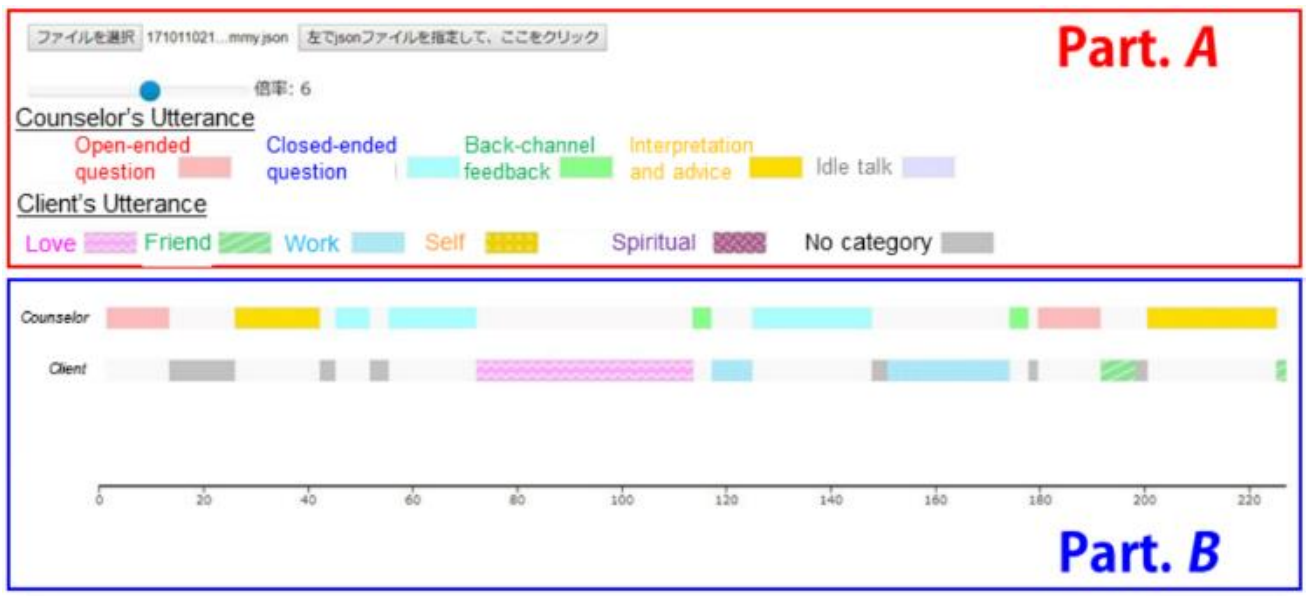

Fig. 26 – Gantt Chart metaphor generated in SR26 for a conversation between a psychology counselor and the client. We note that the categories of the counselors are different from the clients. Once each talk-turn is categorized, a Gantt Chart metaphor summarizing the conversation is produced. The x-axis represents the cumulative number of words (the source language is Japanese).

# Appendix 3 – Search terms queried in the databases

Table 6
Details of search terms, database and query dates

| Query Date | Database | Query |
| --- | --- | --- |
| 18-01-18 | IEEE | (((((conversation) OR dialogue) OR chat) OR messag*) AND "IEEE Terms":Visualization) |
| 18-01-18 | ACM | acmdlCCS:(+Visualization) AND (Conversation Dialogue Chat Messag*) |
| 18-01-18 | Scopus | ( ALL ( "conversation" AND "chat" AND "dialogue" AND "messag*" ABS-KEY ( visualization ) ) AND PUBYEAR > 1997 |
| 18-01-18 | Web Of Science | (TI=(conversation OR chat OR dialogue OR messag*)) AND (TI=(Visuali Indexes=SCI-EXPANDED, SSCI, A&HCI, CPCI-S, CPCI-SSH, ESCI, CCR-EXP Timespan=1998-2018 |
| 18-01-18 | ACM | "query": { recordAbstract:(Visualization Visualisation) AND (Conversat Chat Messag* -chatbot) AND acmdlCCS:(+"Human-centered computin<br><br>"filter": {"publicationYear":{ "gte":1998 }},<br>{owners.owner=GUIDE} |
| 25-01-18 | ACM | acmdlTitle:(+Visuali* Conversation Dialogue Chat Messag* -chatbot -t interface -impaired -blind -haptic -predict* -tour -gaze -multimodal -m human-machine -uml -xml -java -oracle -sql -pmp -c# -c++ -c -python - android -mobile -gimp -geospatial -gene -image -photo -face -equatio corporate -government -opengl -optimization -flash -carbon -song -bit lighting -wireless -exam -soa -asp* -enterprise -microsoft -windows -v autodesk -shadow* -castle* -jasper* -photoshop -reality -gravit* -tissu sharepoint) AND acmdlCCS:(-database -simulation -biology -genetics - theory -mathematics -retrieval -engineering -image -algorithms -hardw network) |
| 25-01-18 | ACM | acmdlTitle:((visual* map explor*) (theme* topic* text document conv dialog* chat messag*)) AND (+dyad*) |
| 31-01-18 | IEEE | ( "Document Title":visual* OR "Document Title":map OR "Document Title":explor*)<br>AND (dyad*)<br>AND (themes OR topics OR text OR documents OR conversations OR d chats OR messages) |
| 31-01-18 | Scopus | TITLE( "visual*" OR "map" OR "explor*")<br>AND TITLE("theme*" OR "topic*" OR "text" OR "document" OR "conve "dialog*" OR "chat" OR "messag*")<br>AND ALL("dyad*")<br>AND PUBYEAR > 1997 |
| 31-01-18 | Web Of Science | (( "Document Title":visual* OR "Document Title":map OR "Document Title":explor*) AND (dyad*) AND (themes OR topics OR text OR docum conversations OR dialogs OR chats OR messages)) |
| 17-12-18 | IEEE | (((((conversation) OR dialogue) OR chat) OR messag*) AND "IEEE Terms":Visualization)<br>Year = 2018 |